\documentclass[pdftex]{JINST}
\pdfoutput=1

\usepackage[english]{babel}
 
\usepackage{graphicx}
\usepackage{lineno}
\usepackage{wrapfig}
\usepackage{subfigure}
\usepackage{rotating}
\usepackage{amssymb}
\usepackage{mathptmx}
\usepackage{multirow}

\usepackage[T1]{fontenc}
\usepackage[utf8]{inputenc}
\usepackage{babel}
\usepackage[font=small,labelfont=bf]{caption}
\usepackage{footmisc}

\usepackage[numbers]{natbib}
\title{Characterization of the Hamamatsu H12700A-03 and R12699-03 multi-anode photomultiplier tubes}
\author{M. Calvi$^{ab}$, P. Carniti$^{ab}$,  L. Cassina$^{ab}$\thanks{Corresponding author}~, C. Gotti$^{ab}$, M. Maino$^{ab}$, C. Matteuzzi$^{a}$ and G.~Pessina$^{ab}$\\
\llap{$^a$} INFN, Sezione di Milano Bicocca,\\
Piazza della Scienza 3, 20126, Milano, Italy\\
\llap{$^b$} Dipartimento di Fisica G. Occhialini, Universit\`a degli Studi di Milano Bicocca,\\
Piazza della Scienza 3, 20126, Milano, Italy\\
E-mail: \email{lorenzo.cassina@mib.infn.it}}

\abstract{The H12700 is a novel 64-channel 52 $\times$ 52 mm$^2$ square Multi-Anode PhotoMultiplier Tube (MaPMT) produced by Hamamatsu. Its characteristics make this device suitable for high energy physics applications, such as in Ring Imaging Cherenkov (RICH) detectors. Hamamatsu provides the H12700 tube with an embedded socket connecting the anodes to the output pins and including an active voltage divider. A second device version, the R12699, is also available and differs from the former by the absence of the socket. This paper describes a complete characterization of both models, starting from the standard operating parameters (single photon spectra, average gain, anode uniformity and dark current value), investigating in detail the cross-talk effect among neighbouring pixels and considering the behaviour in critical environment conditions, such as in presence of a static magnetic field up to 100 Gauss, at different operating temperatures and after long exposure to intense light.}


\begin{document}
%%%%%%%%%%%%%%%%%%%%%%%%%%%%%%%%%%%%%%%%%%%%%%%%%%%%%%%%%%%%%%%%%%%%%%%%%%%%%%%%%%%%%%%%%%%%%%%%%%%%%%%%%%%%%

\section{Introduction}

\begin{figure}[h!]
	\centering
		\begin{minipage}[t]{.4825\textwidth}
			\includegraphics[width=1\textwidth]{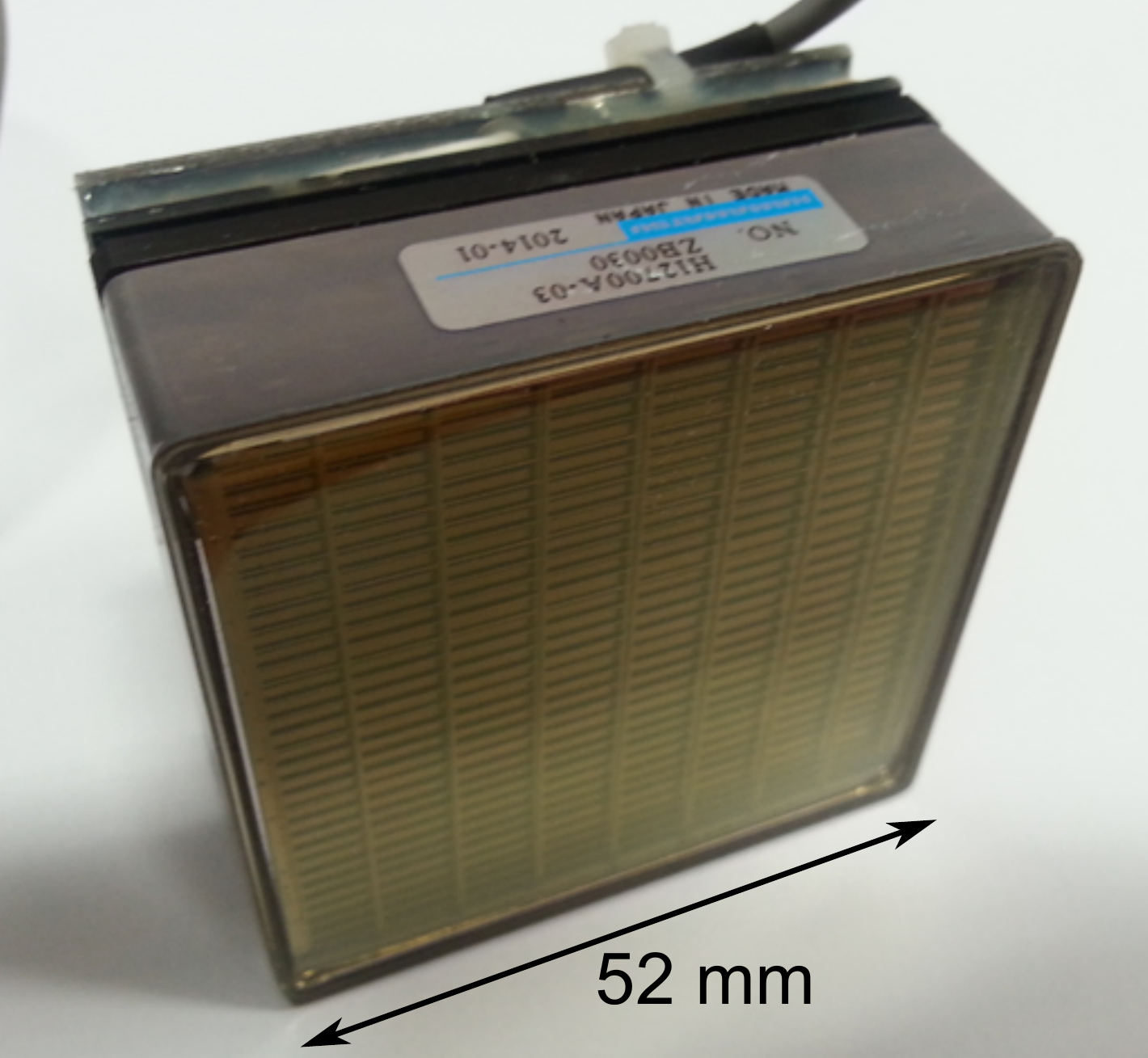}
			\caption{Front view of the Hamamatsu H12700 (serial code: ZB0030, produced in January 2014).}
			\label{fig:PMTFronte}
		\end{minipage}%
	\hspace{5mm}%
		\begin{minipage}[t]{.4825\textwidth}
			\includegraphics[width=1\textwidth]{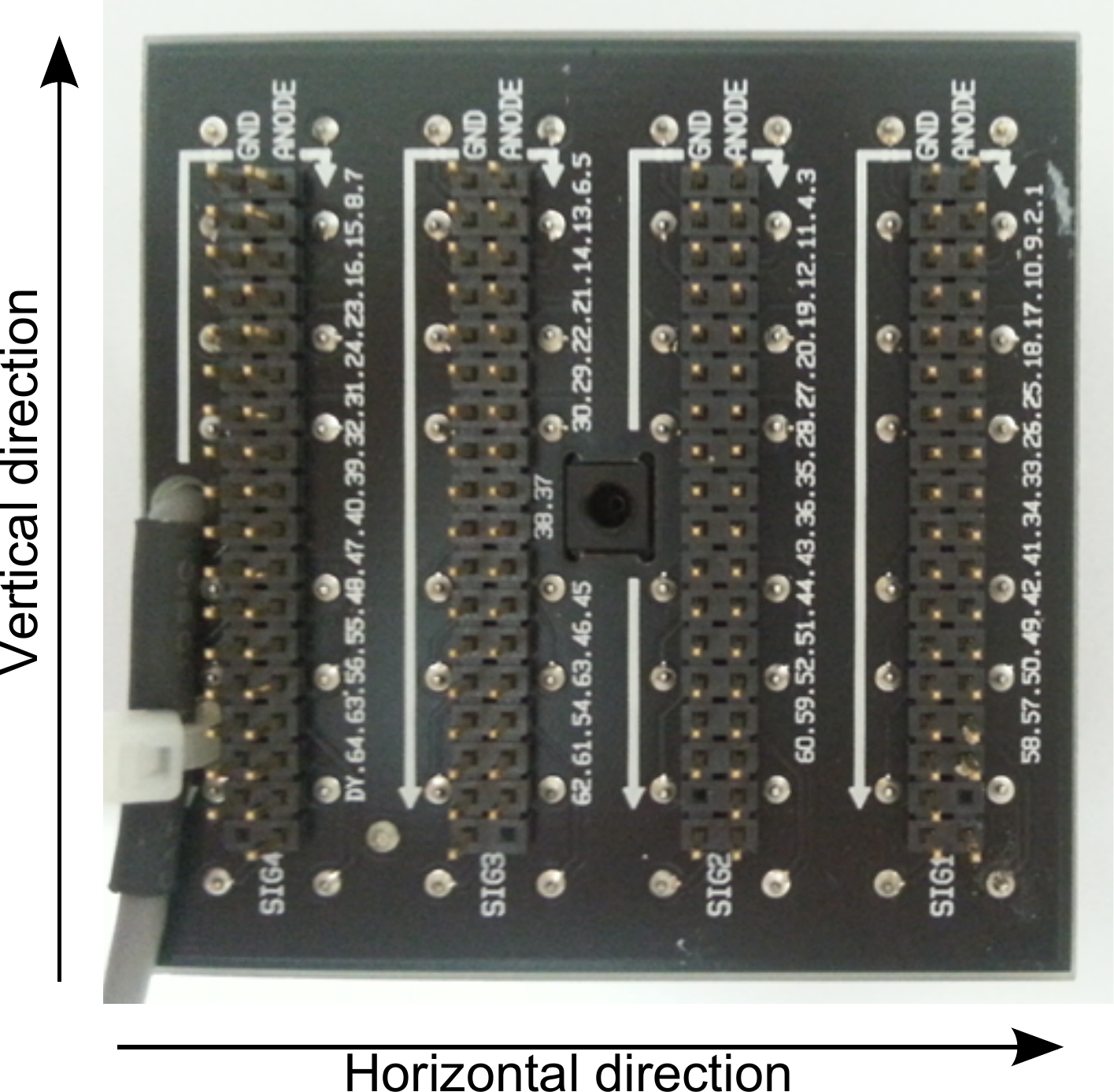}
			\caption{Back side of the H12700. The output pins of the socket embedded with the MaPMT are visible. The convention used to indicate the horizontal and vertical directions is shown.}
			\label{fig:PMTZoccolo}
		\end{minipage}
\end{figure}

The H12700A-03-M64 Multi-Anode PhotoMultiplier Tube is a novel 64-channel, 52 $\times$ 52~mm$^2$  square pixelated device produced by Hamamatsu (fig.\ref{fig:PMTFronte}). It consists of a 8$\times$8 pixel matrix (6$\times$6~mm$^2$ each) able to detect single photons, amplifying the signal through a 10-stage dynode chain. The H12700 MaPMT can be equipped with a borosilicate or UV glass entrance window coupled with a bialkali photocathode so that a spectral response ranging between \hbox{300 - 650 nm} or  \hbox{185 - 650 nm} respectively is obtained, with a maximum quantum efficiency of  $\sim$33\%\footnote{\label{HamamatsuData} Reference value provided by the manufacturer.} at about 350 nm.  A socket, provided by Hamamatsu as standard and embedded with the H12700 MaPMT (fig.\ref{fig:PMTZoccolo}), connects the anodes to the output pins and includes the HV divider. Note that all the measurements described below were acquired using such an embedded voltage distribution ratio: \hbox{2-1-1-...-1-1-0.5}\footnote{~Standard configuration suggested by the manufacturer.}. A second tube version is also available, called R12699-03, which has the same structure as the H12700 except for the absence of the socket. In this case, the output pins are connected directly to the anodes and the HV bias of the dynodes was provided by a custom-made passive voltage divider.

Among the multi-channel photodetectors, the H12700 stands out for its large effective active area\footref{HamamatsuData} (48.5 $\times$ 48.5 mm$^2$) while the very small inactive border around the device and the MaPMT square cross-sectional geometry allows for a close packing ratio ($\sim$87\% \footref{HamamatsuData}). These features, together with the nominal low dark counts contribution and the moderate cross-talk between neighbouring pixels, make the H12700 particularly suitable for applications in high energy physics, such as in Ring Imaging Cherenkov detectors (RICH) used for particle identification. 

A possible application is the upgrade of the LHCb \cite{bib2} RICH detector \cite{bib1}. Currently the detector uses hybrid photodetectors (HPDs) equipped with an encapsulated electronics operating at 1~MHz. These must be replaced by photodetectors read-out by an external 40~MHz circuitry \cite{CLARO}, in order to allow the detector to run at a luminosity up to ten times higher than the current value. The baseline tube chosen for the upgraded LHCb RICH detector~\cite{bib3} is the R11265 MaPMT~\cite{NostroArticolo} which ensures an excellent spatial resolution thanks to the small pixel size (2.9$\times$2.9~mm$^2$). Nevertheless, in the peripheral areas of the LHCb RICH-2 detector the spatial resolution is mainly limited by the chromatic dispersion of the refractive index of the Cherenkov medium (CF$_4$), by the uncertainties affecting the reconstructed track position from which the photons are emitted and by the aberration of the focusing mirrors~\cite{bib1}. In those areas, devices  with larger granularity such as the H12700 can be used without affecting the overall PID performance of the system. This choice reduces the number of devices and channels with a consequent lower cost.

Five devices have been tested so far (three H12700 and two R12699 MaPMTs), each equipped with a UV glass entrance window and bialkali photocathode and the outcomes are presented in this paper. In section \ref{sec:StandardMeasurements} the studies of gain variation, anode uniformity, dark current and pixel collection efficiency are presented. The cross-talk signal induced between neighbouring pixels was investigated in depth and the results are presented in section \ref{sec:CrossTak}. Section \ref{sec:Environment} describes the characterization of the tube in the critical environment conditions expected in high energy physics applications such as the temperature dependence of gain and noise, the deterioration of the photodetector performance due to an external static magnetic field and the gain variation caused by the device aging after a long period of intense light exposure.

%%%%%%%%%%%%%%%%%%%%%%%%%%%%%%%%%%%%%%%%%%%%%%%%%%%%%%%%%%%%%%%%%%%%%%%%%%%%%%%%%%%%%%%%%%%%%%%%%%%%%%%%%%%%%

\section{Standard characterization}\label{sec:StandardMeasurements}

\begin{figure}[h!]
	\centering
		\begin{minipage}[t]{.4825\textwidth}
			\includegraphics[width=1\textwidth]{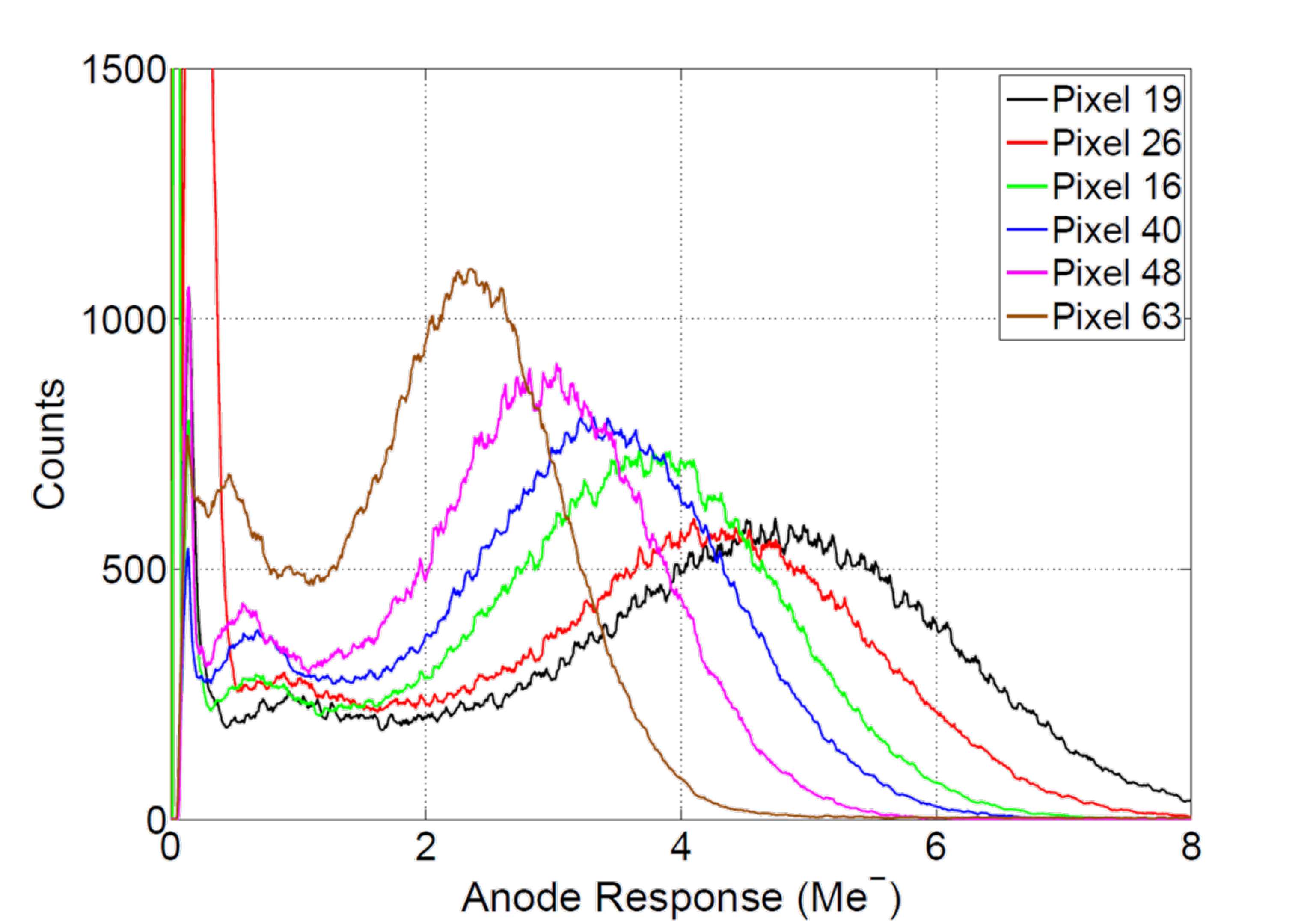}
			\caption{Superposition of single photon spectra acquired on different pixels of the H12700 MaPMT (serial code: ZB0030) biased at 1050~V. Note that pixels 19 and 63 are the ones with the maximum  ($\sim 4.9$~$\mathrm{Me^-}$) and minimum ($\sim 2.4$~$\mathrm{Me^-}$) gain respectively. For the pixel numbering, see fig.\ref{fig:NoiseTable}. }
			\label{fig:SpetraUniformity}
		\end{minipage}%
	\hspace{5mm}%
		\begin{minipage}[t]{.4825\textwidth}
			\includegraphics[width=1\textwidth]{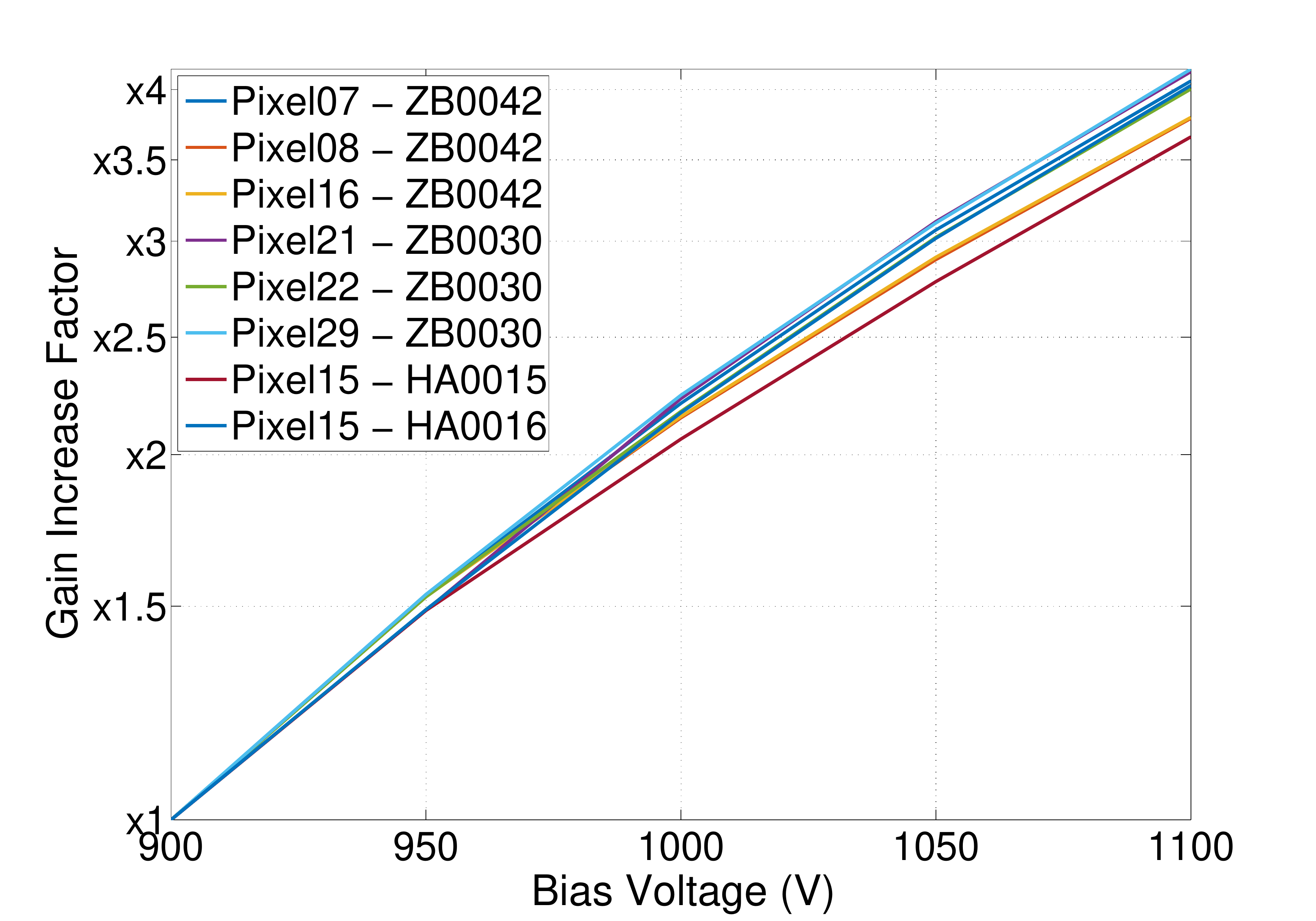}
			\caption{Gain increase factor (in logarithmic scale) measured on several pixels of two H12700 (serial codes: ZB0030 and ZB0042) and R12699 (serial codes: HA0015 and HA0016) MaPMTs as a function of the biasing voltage ranging from 900~V to 1100~V. }
			\label{fig:Gain_Vs_HV}
		\end{minipage}
\end{figure}

\begin{figure}[h!]
	\centering
		\begin{minipage}[t]{.4825\textwidth}
		\centering
			\includegraphics[height=0.7\textwidth]{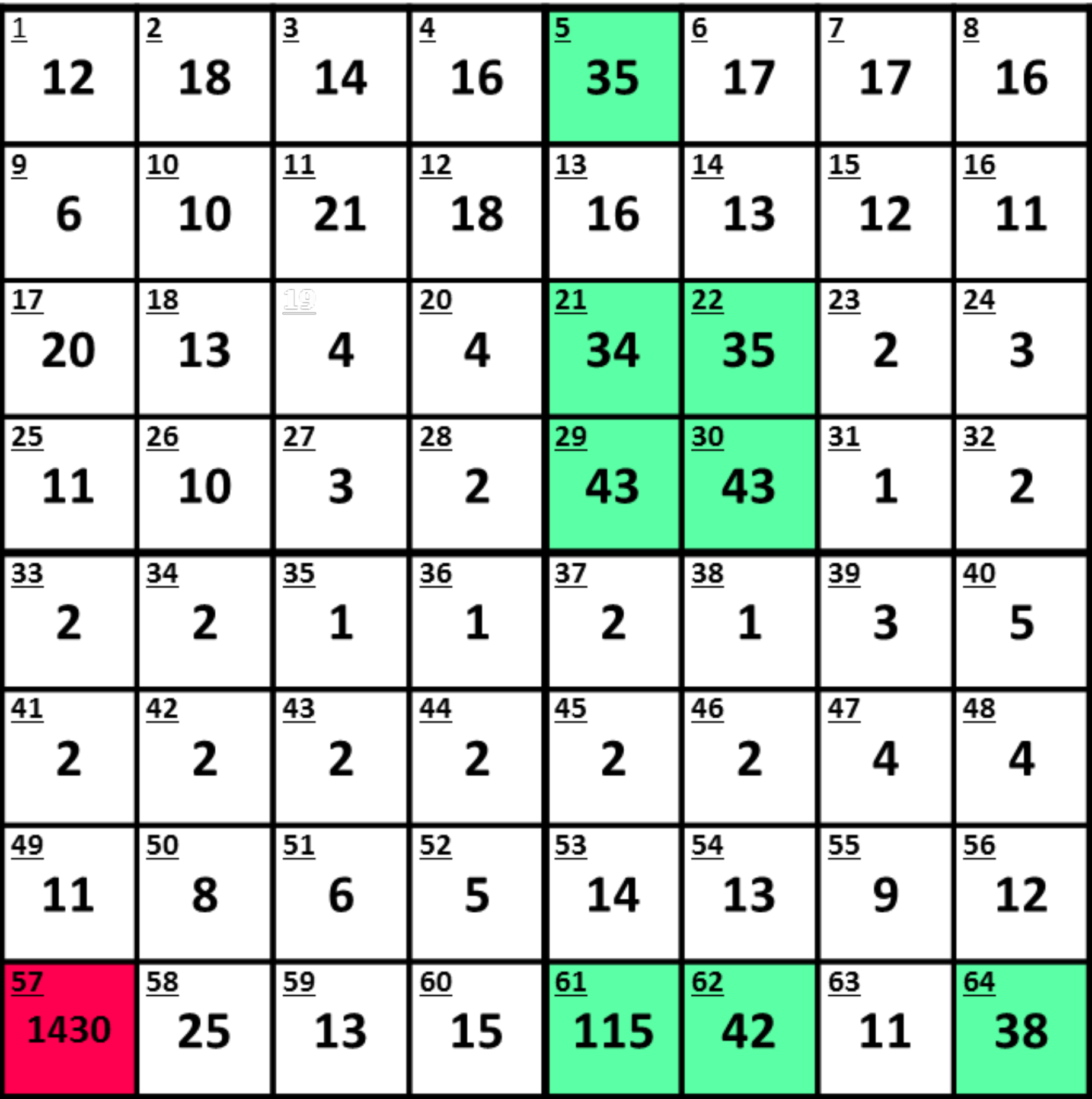}
			\caption{Dark event rate (in Hz) measured for the H12700 MaPMT pixels over a 1 $\mathrm{Me^-}$ amplitude threshold (H12700 MaPMT, serial code: ZB0030).}
			\label{fig:NoiseTable}
		\end{minipage}%
	\hspace{5mm}%
		\begin{minipage}[t]{.4825\textwidth}
			\includegraphics[width=1\textwidth]{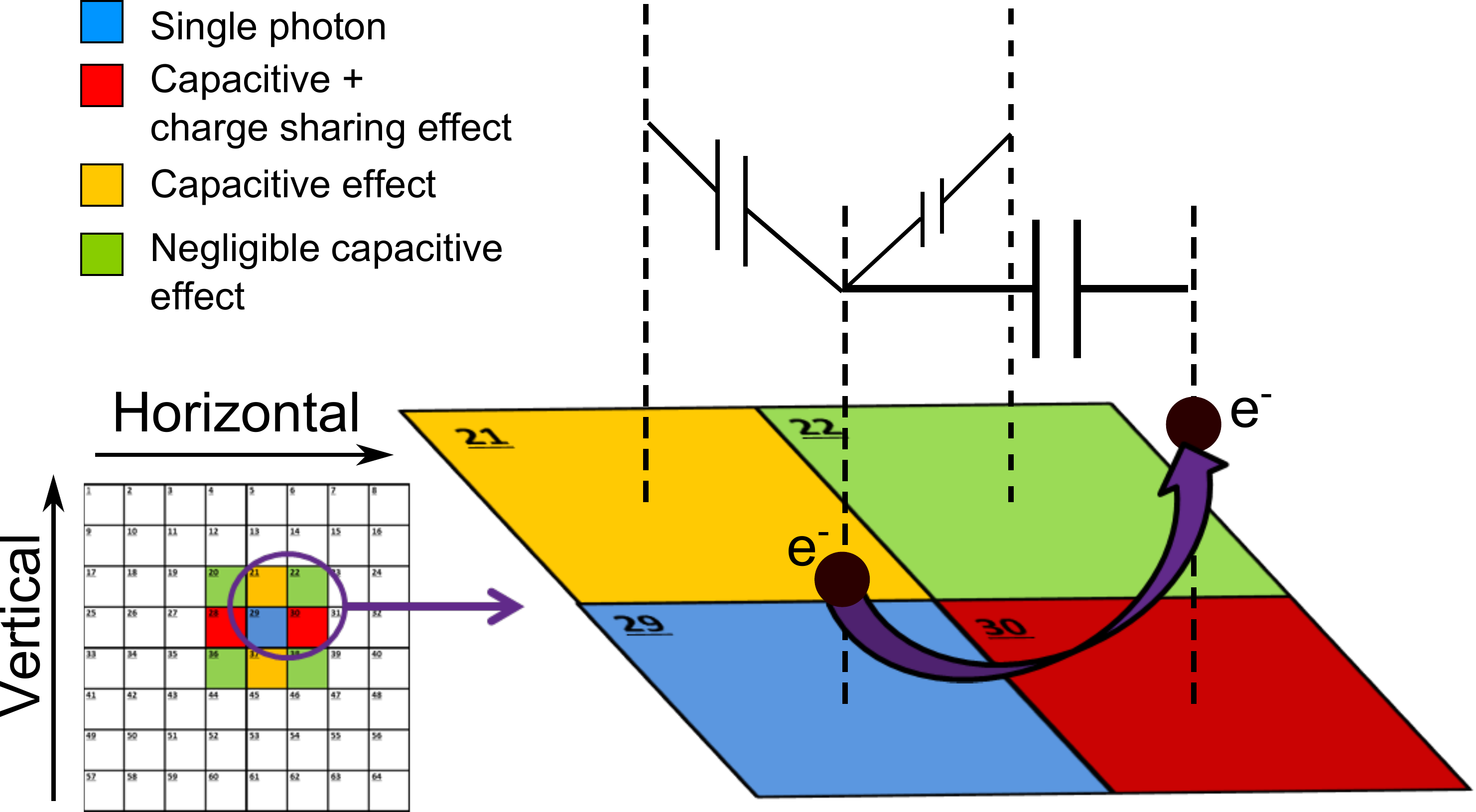}
			\caption{Schematic representation of the different sources of cross-talk usually affecting a MaPTM. This scenario is compatible with the results of the cross-talk measurements.}
			\label{fig:schemino}
		\end{minipage}
\end{figure}

This section is devoted to the measurements of the basic parameters characterizing the operation of the H12700 and R12699 MaPMTs in single photon detection regime. The following results were achieved by illuminating uniformly the device under test with a commercial blue LED biased with a very small current so that it produced only few tens of photons per second. A photon reaching the photocathode and converted into one photoelectron via photoelectric effect starts the 10-stages multiplication process on the dynodes so that a final charge of the order of few Me$^-$ is collected at the anode. This signal is integrated by a charge sensitive preamplifier (standard current integrator circuit based on an AD9631 voltage feedback operational amplifier, by Analog Devices) producing an output voltage signal proportional to the total input charge, recorded by a CAEN Desktop Digitizer DT5720\footnote{~The DT5720 is a 4 channels 12 bit 250 MS/s Desktop Waveform Digitizer with 2 Vpp single ended input. The DC offset is adjustable via a 16 bits DAC on each channel in the $\pm$ 1V range.}. Unless otherwise noted, the tests were performed at room temperature and biasing the device at HV=1050 V using the voltage divider recommended by Hamamatsu (voltage ratio distribution: 2-1-1-...-1-1-0.5).
In fig.\ref{fig:SpetraUniformity} the superposition of typical single photon spectra from different pixels of a H12700 MaPMT (serial code: ZB0030) are shown. All the pixels, except pixel 57\footnote{~Pixels 57 shows a dark counts rate higher than 1 kHz, so that the single photon peak could not be studied with the described setup (maximum single photon production rate of $\sim 1$ kHz per pixel). As declared by Hamamatsu, this anomalous value is due to a manufacturing defect. No other tested device shows analogous behaviour.} (see fig.\ref{fig:NoiseTable}), are able to detect single photon signals with the single photon peak well resolved with respect to the pedestal. The low amplitude peaks just above the pedestal (centered at $\sim 20\%$ with respect to the single photon peak position) could be interpreted as a non ideal electron paths along the multiplication chain, such as photoelectrons skipping the first dynode, or photons that are converted into electrons at the first dynode stage. As will be described in section  \ref{sec:Environment}, this peak becomes higher in the presence of an external magnetic field which affects the ideal trajectory of the secondary electrons moving from one dynode to the next. The anode uniformity can be estimated from fig.\ref{fig:SpetraUniformity} as the ratio between the single photon peak position in each pixel with respect to the pixel with the maximum gain. The typical gain spread among pixels amounts to a factor 2-2.5 at most, in agreement with the uniformity tables provided by Hamamatsu. At 1050 V, the H12700 under investigation provides a typical gain of $\sim 3.5$~$\mathrm{Me^-}$/photon. Similar gains and spreads were observed also in the other tested devices.

By measuring the single photon peak position at different biasing voltages, it is possible to determine how the gain of the photodetector scales as a function of the HV. Figure \ref{fig:Gain_Vs_HV} shows the increase of the gain with respect to the HV value ranging from 900 V to 1100 V for several pixels of the four MAPMTs. It is observed that the gain at HV=1100 V is $\sim$3.5-4 times larger than that at 900 V. These scaling values do not depend on the MaPMT model or on the pixel absolute gain, as expected.

The dark current, i.e. the charge collected at the anodes due to spurious electron emission from the photocathode or the dynodes, represents an important figure of merit of the MaPMT. The dark current is usually quoted as the overall DC anode current measured keeping the MaPMT in the dark. For the studied MaPMTs the dark current at room temperature ranges from $\sim 0.5$~nA to $\sim 3.5$~nA, values lower than the typical value declared by the manufacturer (6~nA). 

In case were the photodetector used in applications which require single photon sensitivity in all the pixels, it is interesting to measure the rate of spurious counts triggered in each channel above a defined threshold. Thus, the MaPMT dark current was also measured at room temperature by recording the rate of events with an amplitude larger than $1$~$\mathrm{Me^-}$ (average position of the valley between the pedestal and the single photon peak, as visible in the spectra in fig.\ref{fig:SpetraUniformity}).  Figure \ref{fig:NoiseTable} shows the dark event rate measured for all the pixels of a tested device. The dark event rate turned out to be really low ($\leq 15$ Hz per pixel or, equivalently, $\leq 40$ $\mathrm{Hz/cm^2}$) for most of the pixels. Few of them have a larger dark current (up to several tens of Hz) and pixel 57 is the only one with an extremely high dark event rate ($\sim 1.4$ kHz).

%%%%%%%%%%%%%%%%%%%%%%%%%%%%%%%%%%%%%%%%%%%%%%%%%%%%%%%%%%%%%%%%%%%%%%%%%%%%%%%%%%%%%%%%%%%%%%%%%%%%%%%%%%%%%

\section{Cross-talk and charge sharing}\label{sec:CrossTak}
A photon hitting a pixel induces a signal in the neighbouring pixels so that they could trigger an event despite being in the dark. This undesired effect is called cross-talk and is caused by different mechanisms, as schematically represented in fig.\ref{fig:schemino}. 

A first cross-talk contribution (CT$_{bias}$) is due to the signal injected through the voltage divider biasing the dynodes unless properly filtered. A second contribution comes from the stray capacitance between neighbouring pixels which results in an AC coupling between the anode pins. This contribution (CT$_{coupling}$) causes not only spurious signals, but it also reduces the single photon signal amplitude at high frequency, as part of the charge deviates from the original path. Note that, in addition to the intrinsic contribution of the photodetector, the CT$_{coupling}$ also depends on the possible stray couplings between the electrical paths driving the signal from the input nodes to the read-out electronics. This suggests that not only particular care must be applied in the design and layout of the front-end circuit, but also that the socket embedded with the H12700 MaPMT could introduce an additional contribution, as it will be shown later. The waveform of the signals due to CT$_{bias}$ and CT$_{coupling}$ is expected to have a bipolar shape. Furthermore, the charge sharing can be considered as a cross-talk source (CT$_{sharing}$), since it causes effects similar to CT$_{coupling}$, but with a different signal shape. The charge sharing is a statistical process which takes place whenever an electron deviates from its path during multiplication, starting a new avalanche in the neigbouring channel. Note that CT$_{sharing}$ gives a negligible contribution unless it starts from the first dynode stages; if the charges are shared only at the the first dynode, then the amplitude of the induced signal is expected to be independent of the gain of the reference pixel. Otherwise, if the charge sharing occurs at few initial dynodes, then the amplitude of the induced signal is proportional to the gain of the inducing pixel. Anyway, what matters for our investigations is that, differently to the previous contributions, the signals due to CT$_{sharing}$ have the same shape as the main signal, so that they can be discriminated from that due to CT$_{coupling}$. 

The cross-talk effect is estimated by measuring the cross-talk ratio distribution. The cross-talk ratio is the cross-talk signal amplitude divided by the corresponding source signal amplitude, both calculated as the difference between the maximum of each signal and its baseline. Since, independently on the sources, the cross-talk signal amplitude is expected to be proportional to the inducing signal amplitude, this distribution should exhibit peaks centered on the values characteristic of the cross-talk source.
These measurements were performed illuminating the whole MaPMT with a blue LED operating in single photon regime and acquiring a cluster of four contiguous pixels simultaneously. The cross-talk signals were discriminated off-line considering the coincidences with single photons hitting the adjacent pixel with signal amplitude larger than $100$ $\mathrm{ke^-}$. A pulse shape analysis is also performed in order to discriminate between unipolar (CT$_{sharing}$ contribution) and bipolar (CT$_{bias}$ and CT$_{coupling}$ contributions) signals. Note that the front-end preamplifier was kept as close as possible to the device under test (pins connected with $\leq10$ cm widely separated unshielded copper wires), minimizing the loss of the signals on the stray capacitance of the wires connecting the anodes and the read-out system. 
In the following, a pair of side-by-side pixels is defined as \textit{horizontally oriented} if the numbers of the pixels are consecutive (see fig.\ref{fig:schemino}) and consequently comes the definition of \textit{vertically oriented} pixels. 

\begin{figure}[h!]
	\centering
		\subfigure[][]{\includegraphics[width=.4825\textwidth]{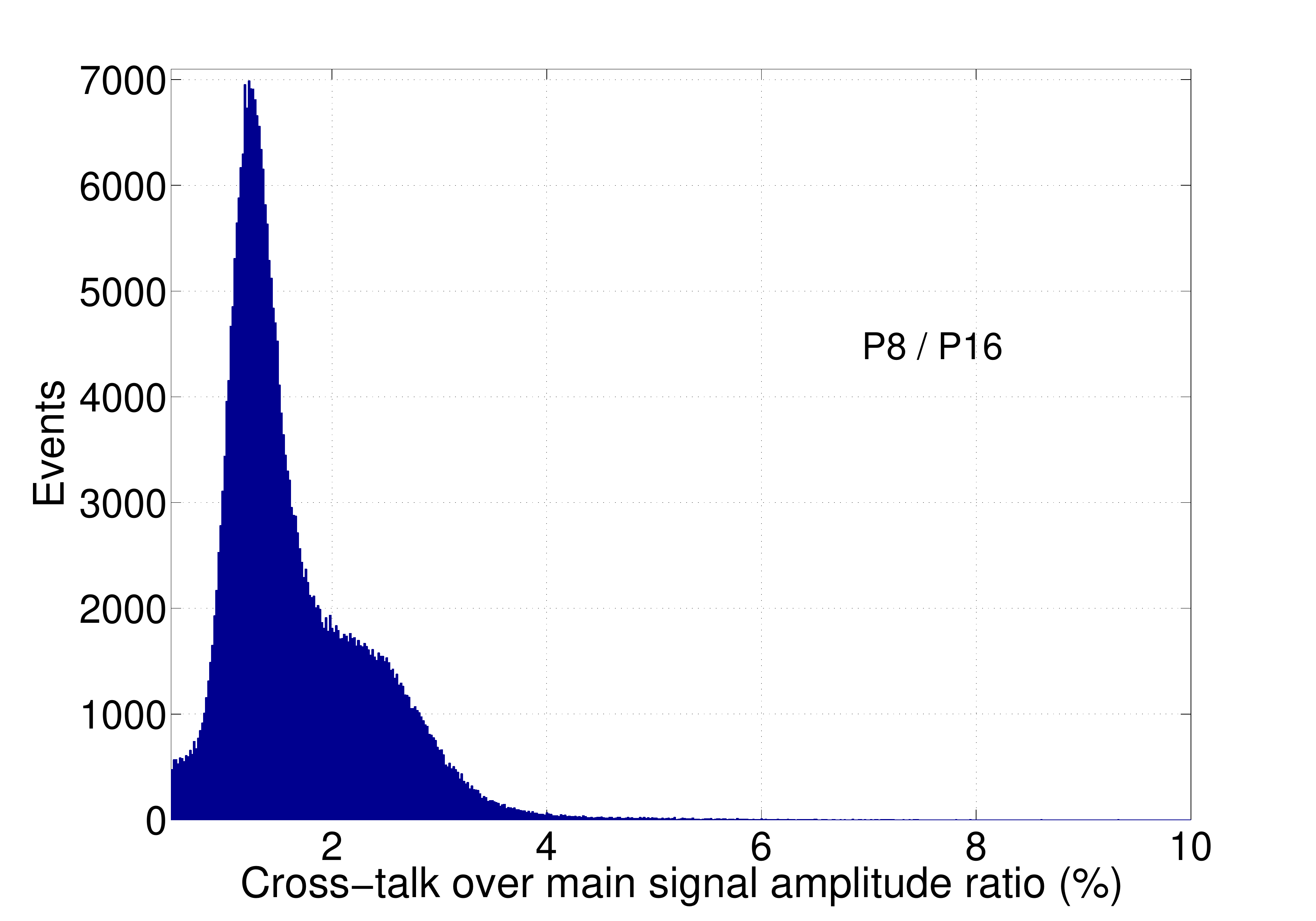}}
		\hspace{1mm}%
		\subfigure[][]{\includegraphics[width=.4825\textwidth]{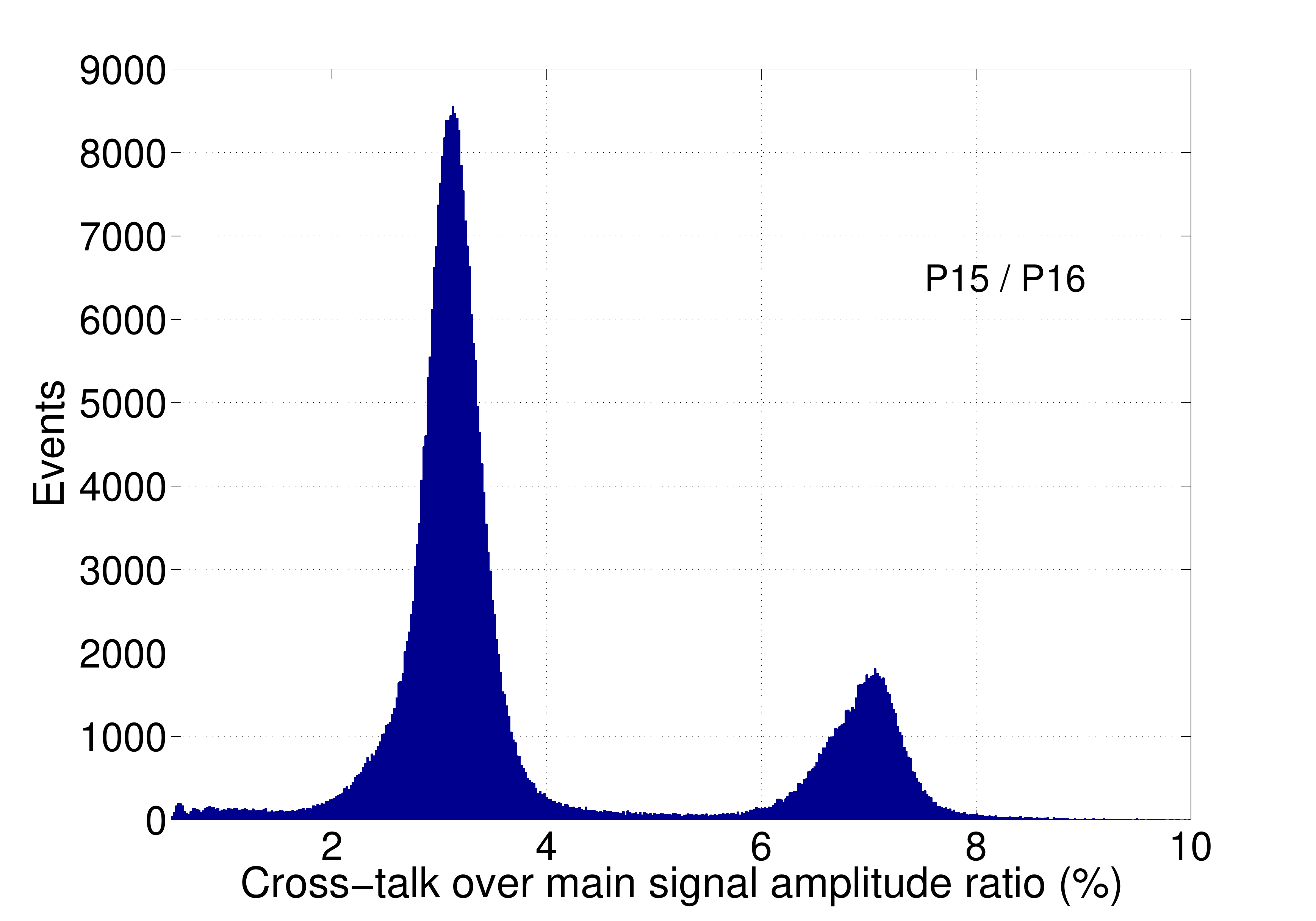}}
		\caption{Cross-talk amplitude distribution of a Hamamatsu H12700 (serial code: ZB0030), obtained illuminating pixel 16 with single photons and acquiring the cross-talk signals on pixels 8 (\textit{a}) and 15 (\textit{b}).}\label{Fig:Cross1}
\end{figure}

\begin{figure}[h!]
	\centering
		\subfigure[][]{\includegraphics[width=.4825\textwidth]{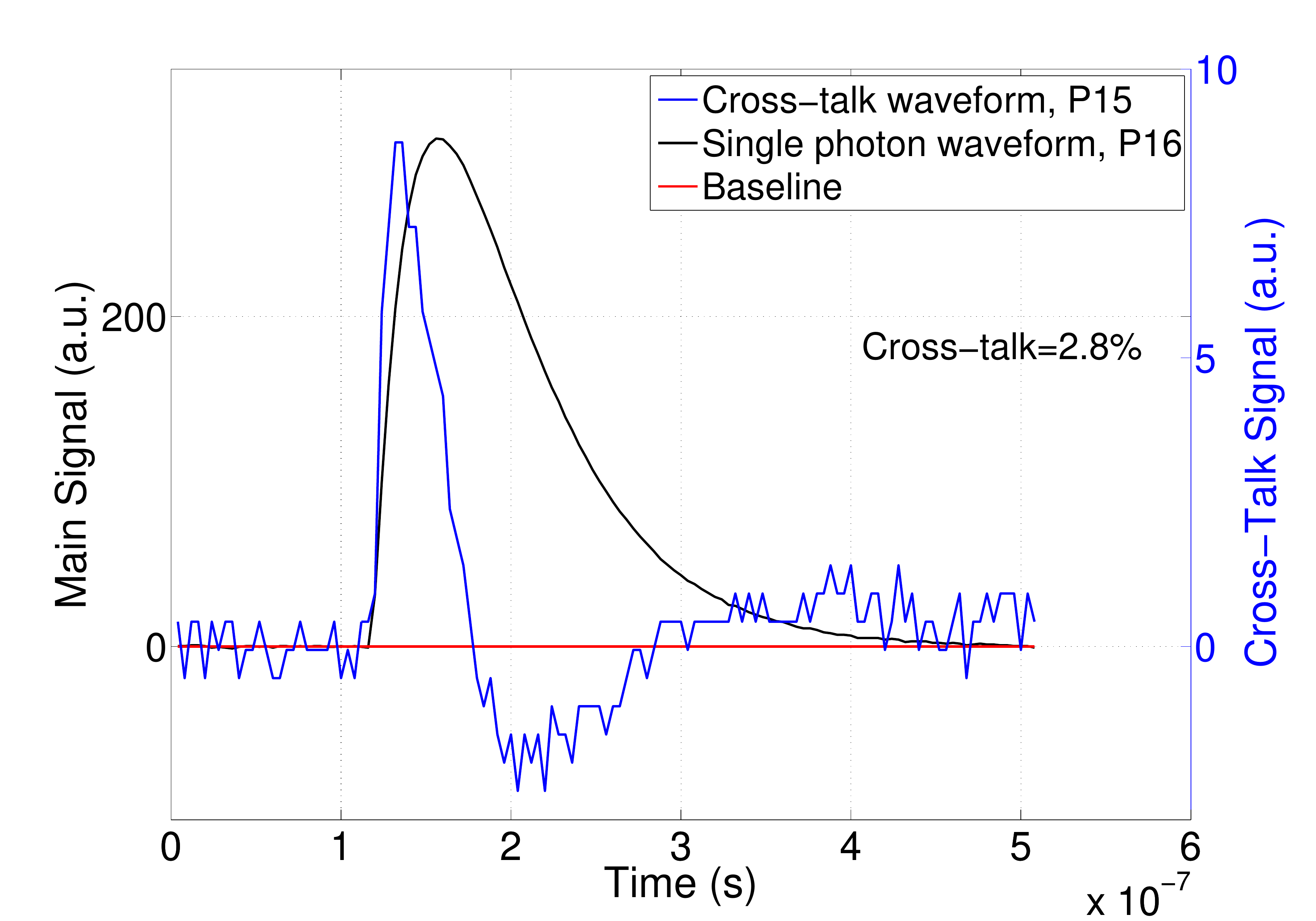}}
		\hspace{1mm}%
		\subfigure[][]{\includegraphics[width=.4825\textwidth]{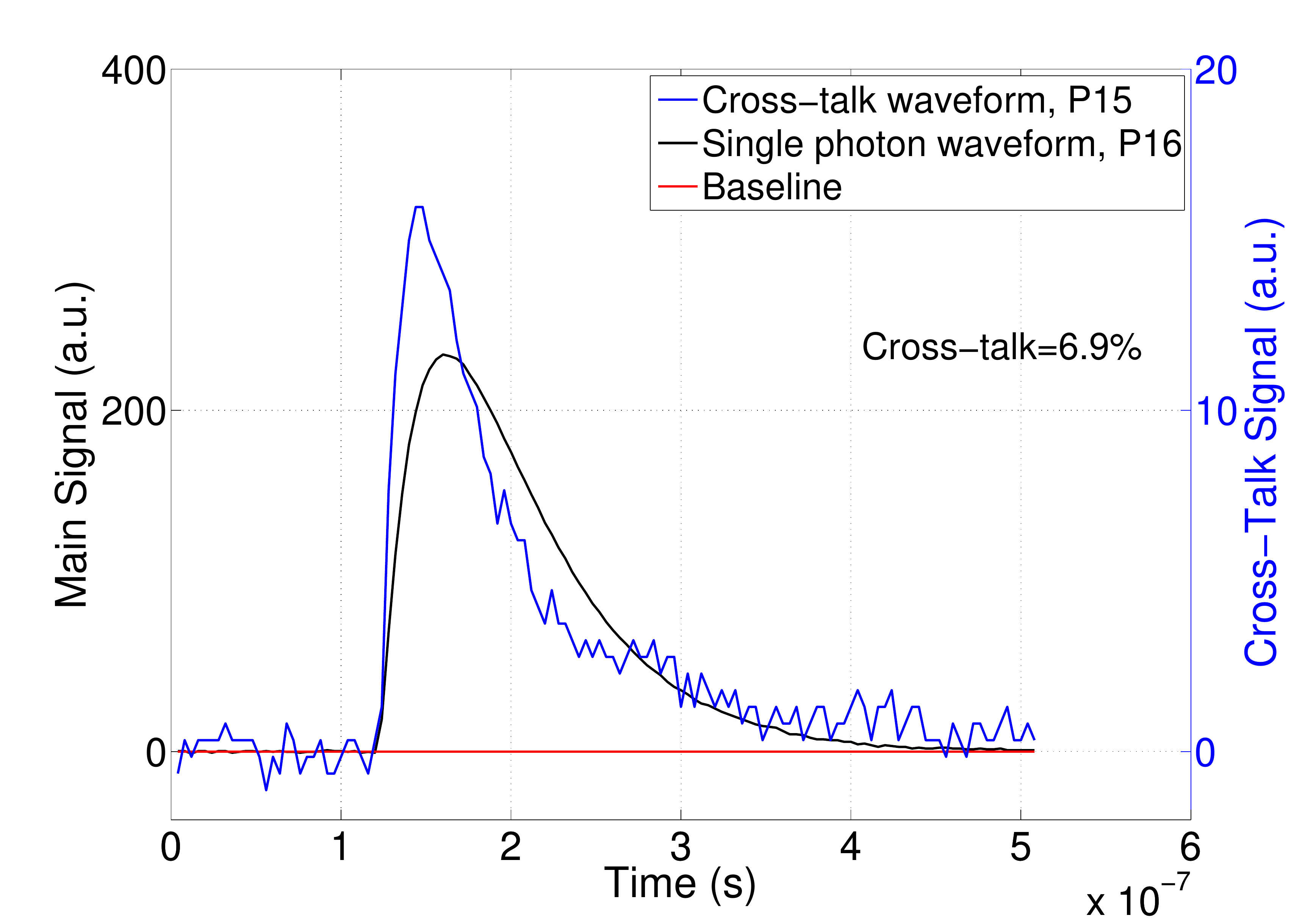}}
		\caption{The typical cross-talk waveforms acquired on pixel 15 induced by a single photon event occurred on the adjacent pixel 16 (horizontal direction). On (a), the typical bipolar shape of the $\sim 3$\% cross-talk events. On (b), the single polarity cross-talk signal. Note that the main signal (black line) and the cross-talk (blue line) are plotted in two different scales.}\label{Fig:CrossWaveforms}
\end{figure}

Figure \ref{Fig:Cross1} shows the cross-talk ratio distribution measured by illuminating pixel 16 with single photons and acquiring the induced signals on pixels 8 (\textit{a}, vertical direction) and 15 (\textit{b}, horizontal direction).
In fig.\ref{Fig:Cross1}.a, an asymmetric peak centered at $\sim$1.3\% and extending up to $\sim$3\% is visible on pixel 8. The behaviour is different for the induced signal on pixel 15 (fig.\ref{Fig:Cross1}.b, horizontal direction). %Both distributions are the result of a single photon event triggered on pixel 16.
In the horizontal direction, two peaks are clearly resolved, one located at around 3\% and a second peak at about 7\%. The signals which populate the two regions have different waveforms, as shown in fig.\ref{Fig:CrossWaveforms}. The waveforms of the $\sim 3$\% cross-talk events (fig.\ref{Fig:Cross1}.a) have a bipolar shape, while the ones corresponding to $\sim 7$\% cross-talk events (fig.\ref{Fig:Cross1}.b) have a single polarity since they always stay above the baseline level\footnote{~As far as the cross-talk ratio distribution of pixel 8 is concerned, fig.\ref{Fig:Cross1}.a, a prevalence of unipolar shaped waveforms can be seen in the asymmetric high-amplitude ratio tail with respect to the bipolar shaped signals mostly populating the main peak. However, the two contributions are not easily separated.}. 

\begin{figure}[h!]
	\centering
		\begin{minipage}[t]{.4825\textwidth}
			\centering
			\includegraphics[width=1\textwidth]{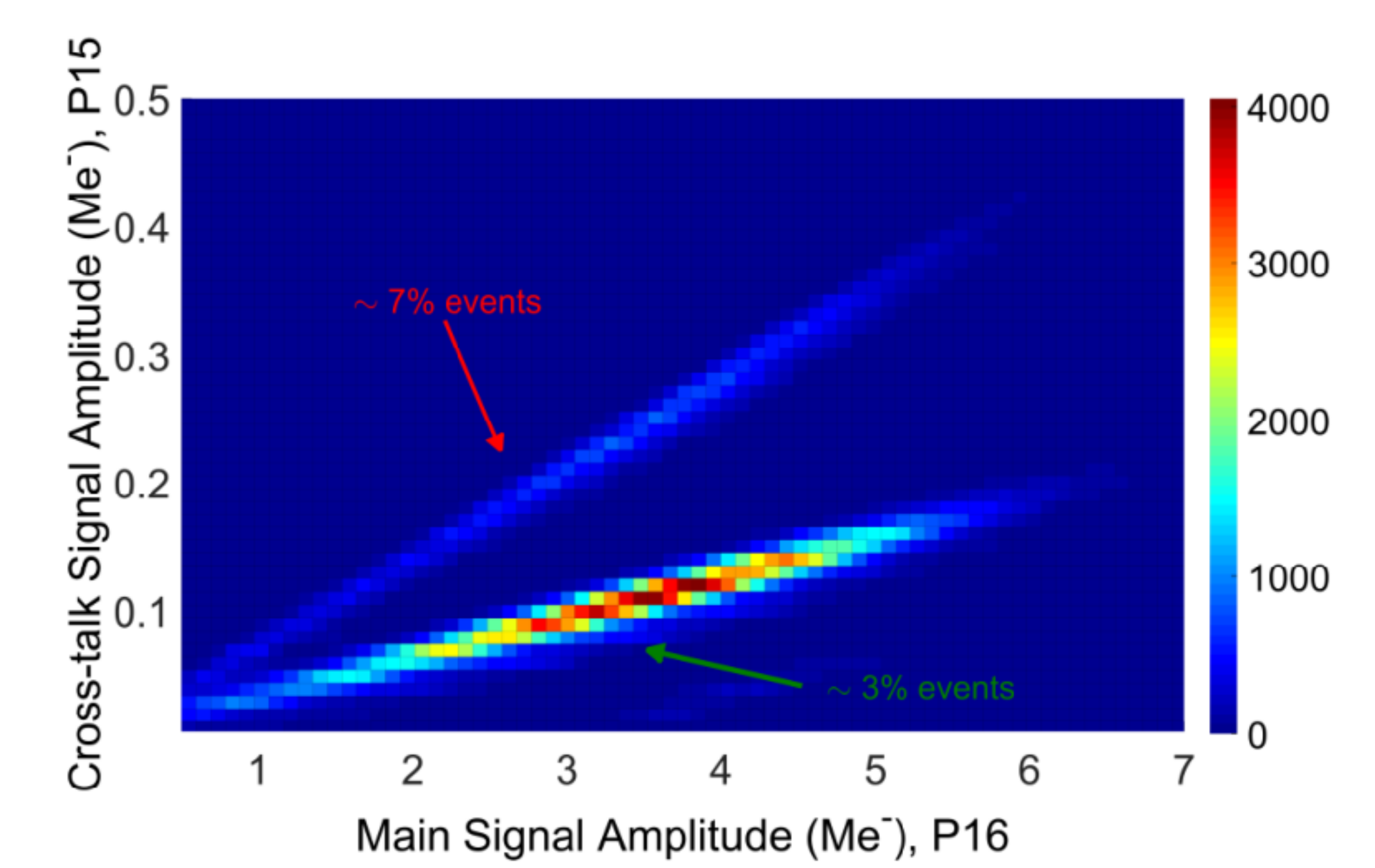}
			\caption{Cross-talk on pixel 15 versus single photon signal amplitude on pixel 16.}
			\label{fig:AmplitudeLin}
		\end{minipage}%
	\hspace{5mm}%
		\begin{minipage}[t]{.4825\textwidth}
			\includegraphics[width=1\textwidth]{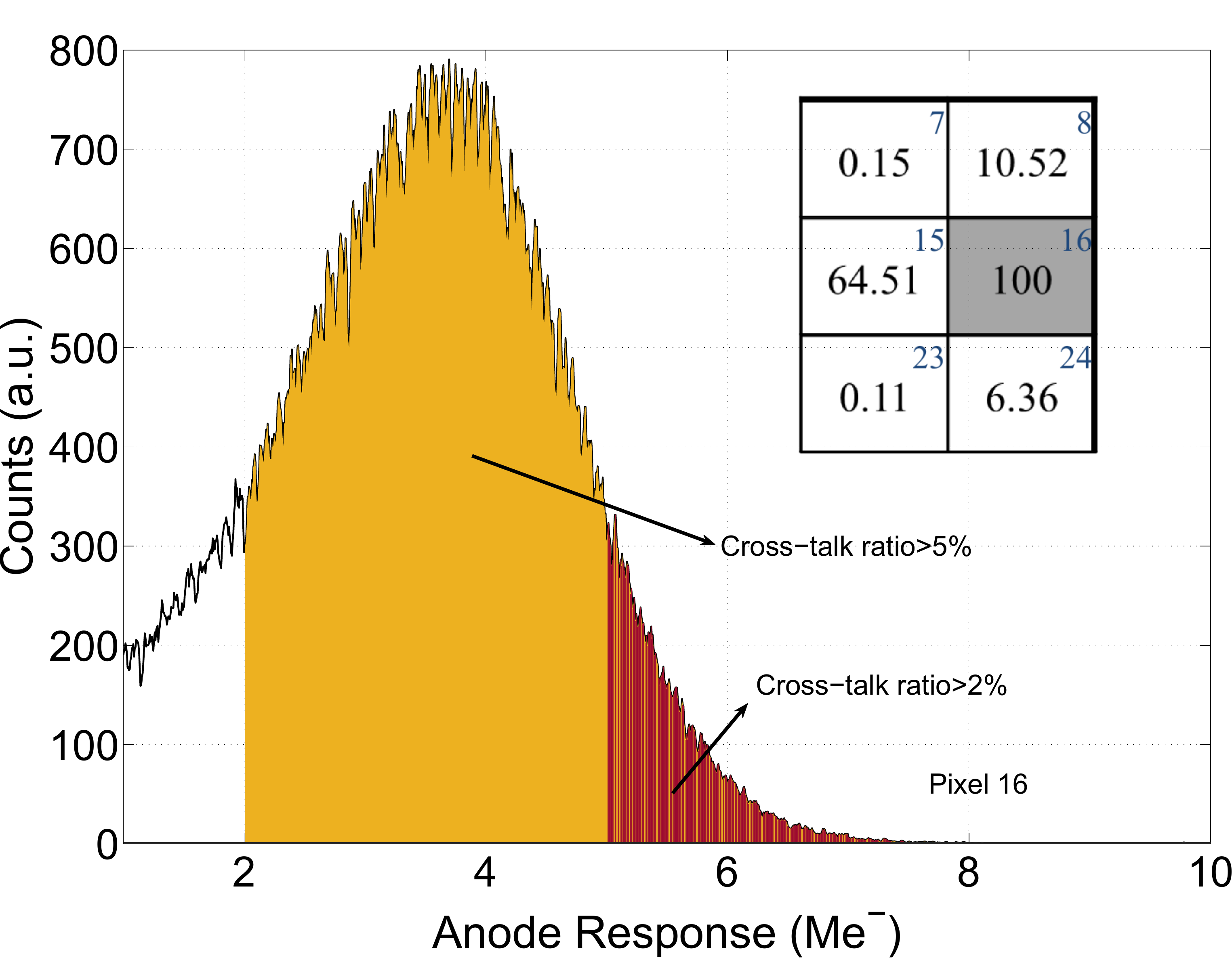}\centering
			\caption{Single photon spectra acquired in pixel 16 over a 1 Me$^-$ threshold. The table shows the probability (in \%) to see a cross-talk signal above 100 ke$^-$ threshold in all the neighbouring channels around around pixel 16. The pixel numbers are reported in blue in the top-right corner.}% (H12700 MaPMT, serial code: ZB0030).}
			\label{fig:CrossTalkProbability}
		\end{minipage}
\end{figure}

Another interesting aspect found in these tests is that both cross-talk behaviours coexist for all amplitudes of the main signal, as shown in fig.\ref{fig:AmplitudeLin}, where the cross-talk amplitude on pixel 15 is plotted as a function of the single photon signal amplitude on pixel 16. Two different linear trends can be seen, with a slope equal to $\sim 3$\% and $\sim 7$\%  respectively. Both lines span from reference signal amplitude of about $\sim$100~$\mathrm{ke^-}$ to several $\mathrm{Me^-}$. Hence, independently of the main signal amplitude, both high-level and low-level cross-talk events are present. 

Figure \ref{fig:CrossTalkProbability} shows the single photon spectrum acquired on the reference pixel 16 over a threshold of 1 $\mathrm{Me^-}$. The probability that these events could induce cross-talk signals larger than 100 $\mathrm{ke^-}$ in the neighbouring pixels is quoted (in percentage unit) in the table shown in the top-right corner. As it can be observed, the probability of inducing cross-talk signal over threshold is of the order of $\sim10\%$ along the vertical direction (pixels 8 and 24) where the CT$_{coupling}$ contribution dominates. Such value is compatible to the one obtained assuming that the mean amplitude of the cross-talk signals due to CT$_{coupling}$ is $\sim2\%$ of the inducing signal amplitude. In this case, only events larger than 5 Me$^-$ (orange area in fig.\ref{fig:CrossTalkProbability}) induce cross-talk over a 100~$\mathrm{ke^-}$ threshold. The same mechanism causes few induced pulses also along diagonal direction (pixels 7 and 23), even if they occur with a much lower probability. 
On the other hand, the CT$_{sharing}$ contribution is responsible of larger cross-talk signals. Assuming that the average cross-talk amplitude ratio on the horizontal direction amounts to $\sim5\%$, then 2 Me$^-$ collected at the pixel 16 anode are enough to induce cross-talk over the selected threshold on pixel 15 (ocher area in fig.\ref{fig:CrossTalkProbability}). This estimation is compatible to the observed probability ($\sim65\%$) that single photon signals larger than 1 $\mathrm{Me^-}$ on pixel 16 cause cross-talk 100~$\mathrm{ke^-}$ threshold on pixel 15.

%\begin{figure}[h!]
%	\centering
%		\begin{minipage}[t]{.4825\textwidth}
%			\includegraphics[width=1\textwidth]{Amplitude3D_P15VsP16_LowResolution.pdf}
%			\caption{Cross-talk on pixel 15 versus single photon signal amplitude on pixel 16.}
%			\label{fig:AmplitudeLin}
%		\end{minipage}%
%	\hspace{5mm}%
%		\begin{minipage}[t]{.4825\textwidth}
%			\includegraphics[width=1\textwidth]{Schemino.pdf}
%			\caption{Schematic representation of the scenario compatible with the cross-talk measurements.}
%			\label{fig:schemino}
%		\end{minipage}
%\end{figure}

In summary, the cross-talk results indicate an unexpected asymmetry between contiguous pixels located on rows or columns. %, a scenario schematically represented in fig.\ref{fig:schemino}. 
Illuminating with single photons a pixel used as reference, the neighbouring pixels along the vertical direction show a cross-talk dominated by CT$_{bias}$ and CT$_{coupling}$ contributions, with an amplitude distribution centered at $\sim1.3 \% -3$\%. Also along the diagonal direction a small coupling between neighbouring pixels is expected. However this turned out to be a negligible effect. 

The cross-talk level between horizontally contiguous pixels deserves more attention. In addition to the previous behaviour, a second gaussian-like peak centered at $\sim4-7\%$ is visible in the cross-talk amplitude distribution. These two peaks are the result of the combination of the CT$_{coupling}$ and CT$_{sharing}$ contributions. Similar outcomes were observed in all the tested pixels of two devices for each MaPMT type (H12700, serial codes: ZB0030 and ZB0042; R12699, serial codes: HA0015 and HA0016). While the double peak structure in the cross-talk ratio distribution was systematically observed along the horizontal direction, the peaks heights differ between pixel pairs and, in first approximation, the probability of charge sharing events ranges from $\sim25\%$ up to $\sim50\%$. This value strongly depends on the position where the inducing photon hits the reference pixel. 

\begin{figure}[ht]
\centering
\includegraphics[width=1\textwidth]{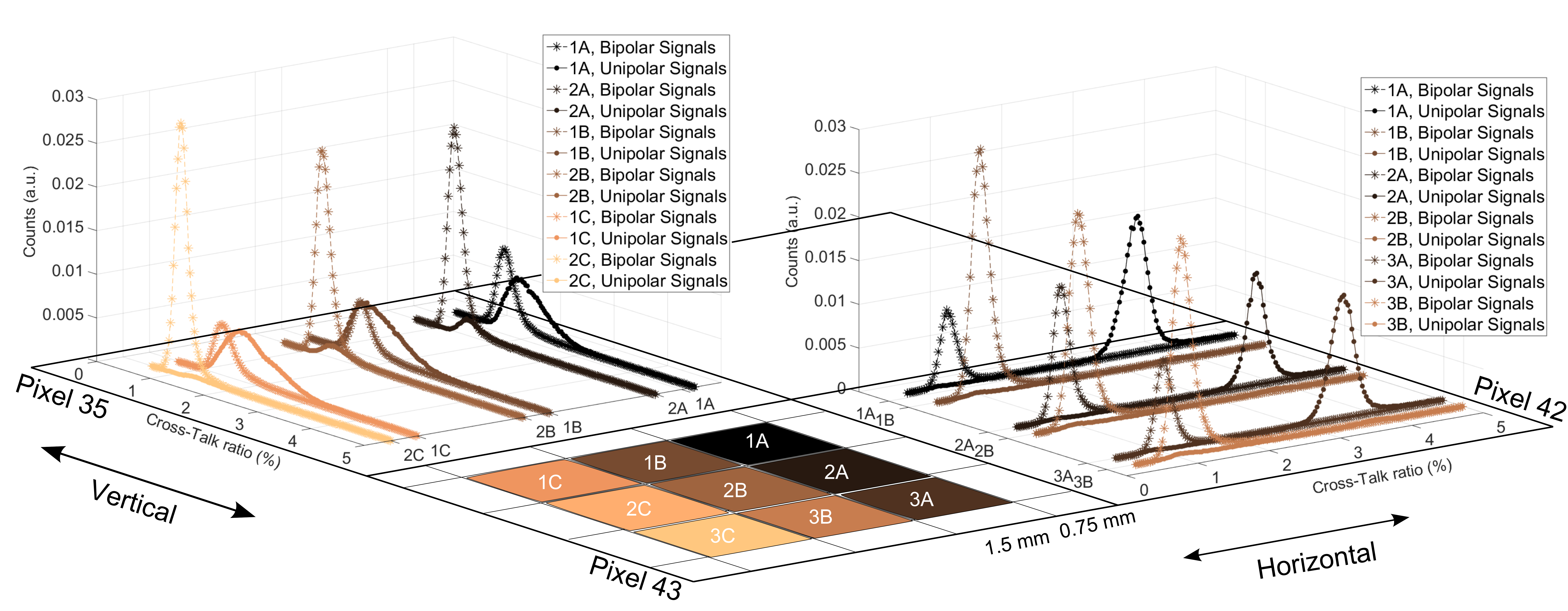}
		\caption{Cross-talk amplitude distribution induced on pixel 42 (horizontal direction) and 35 (vertical direction) by a single photon hitting pixel 43 in different 1.5$\times$1.5 mm$^2$ areas (R12699 MaPMT, serial code: HA0015). Different markers are used for the distributions populated by unipolar (circular) or bipolar (star) shaped cross-talk signals. The plots are colored according to the position where the inducing photon hits the reference pixel.}
		\label{Fig:CrossTalkScan}
\end{figure}

As described above, the uniformity of response of the pixel was studied dividing the pixel area in a $3\times3$ matrix of $\sim1.5\times1.5$ mm$^2$ regions using a mask. Each region can be individually illuminated by means of a 1 mm diameter optical fiber fastened to the mask. 
For example, using pixel 43 as reference, the cross-talk signal is acquired on both horizontal (pixel 42) and vertical (pixel 35) directions. Figure \ref{Fig:CrossTalkScan} shows the obtained results. Along the horizontal direction, two peaks are clearly resolved if the photon hits the A-areas, located at 0.75 mm from the borderline between pixel 43 and 42. In agreement with the previous measurements, the low-level and high-level cross-talk peaks are populated by bipolar (star marker in fig.\ref{Fig:CrossTalkScan}) and unipolar (circular marker) shaped signals respectively. In this case, about $\sim60-70\%$ of single photons induce charge sharing. On the other hand, if the photon hits the B-areas (2.25 mm away from the borderline), a negligible charge sharing is observed since the cross-talk amplitude distribution results in a single low-level gaussian peak. 

In first approximation, assuming the probability of inducing charge sharing to be $65\%$ in the first 2.25 mm from the borderline ($\sim37\%$ of the whole pixel area) and negligible in the other regions of the pixel, $\sim24\%$ of photons would cause charge sharing in the horizontal direction. This coarse estimation is consistent with the results previously shown. 

Similar considerations can be made in the vertical direction (pixel 35). In this case, the two peaks are not well-separated and the charge sharing due to photons hitting the 1-column only causes an asymmetric tail ranging from $\sim1\%$. to $\sim3\%$. In case of uniform illumination in the pixels, the  CT$_{coupling}$ dominates the cross-talk amplitude distribution.

\begin{figure}[h!]
	\centering
		\begin{minipage}[t]{.4825\textwidth}
			\includegraphics[width=1\textwidth]{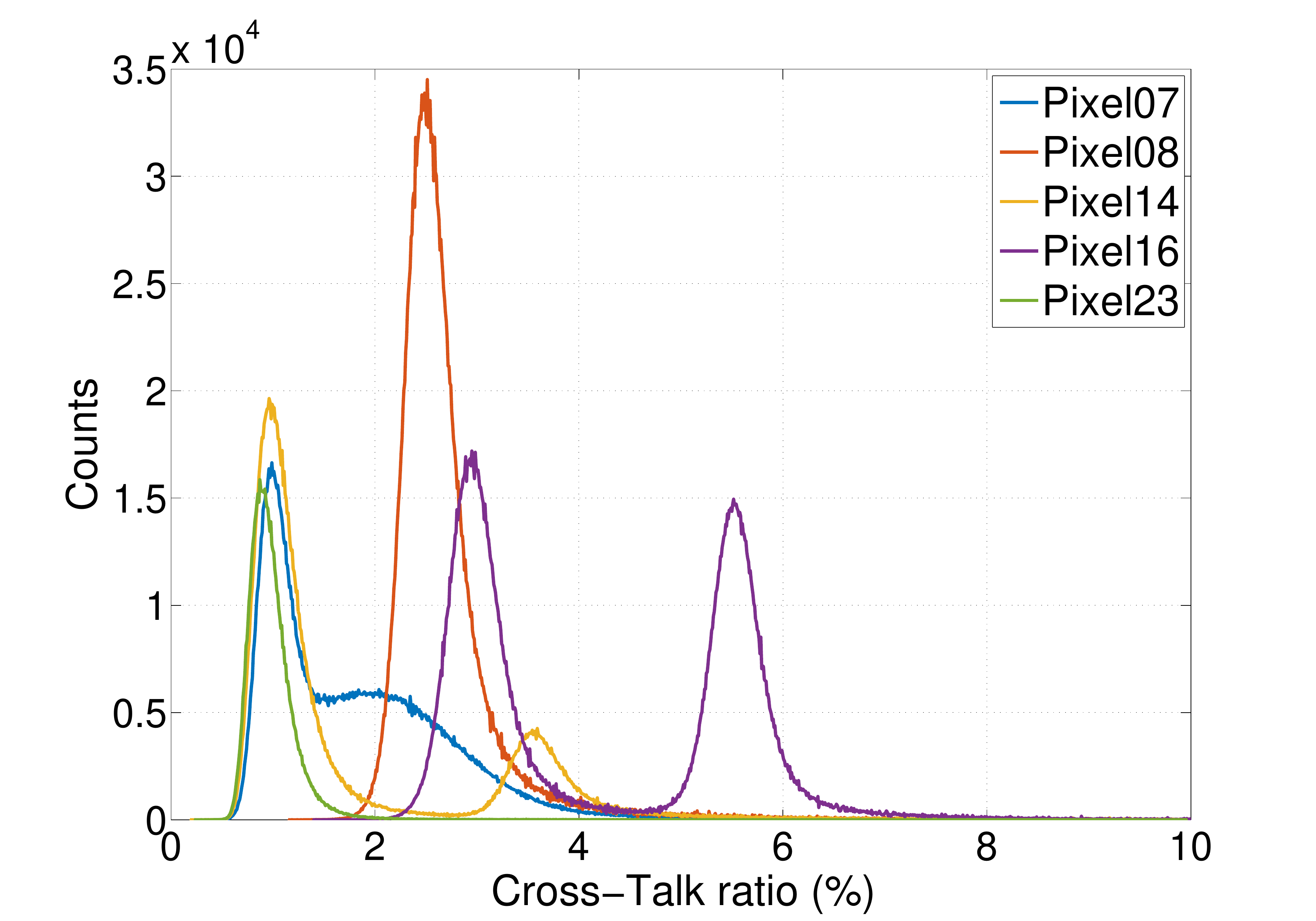}
			\caption{Cross-talk ratio distribution measured on the neighbouring channels of the uniformly illuminated pixel 15 (H12700 MaPMT, serial code: ZB0042). The socket provided by Hamamatsu is connected and operating.}
			\label{fig:CrossTalkSocket}
		\end{minipage}%
	\hspace{5mm}%
		\begin{minipage}[t]{.4825\textwidth}
			\includegraphics[width=1\textwidth]{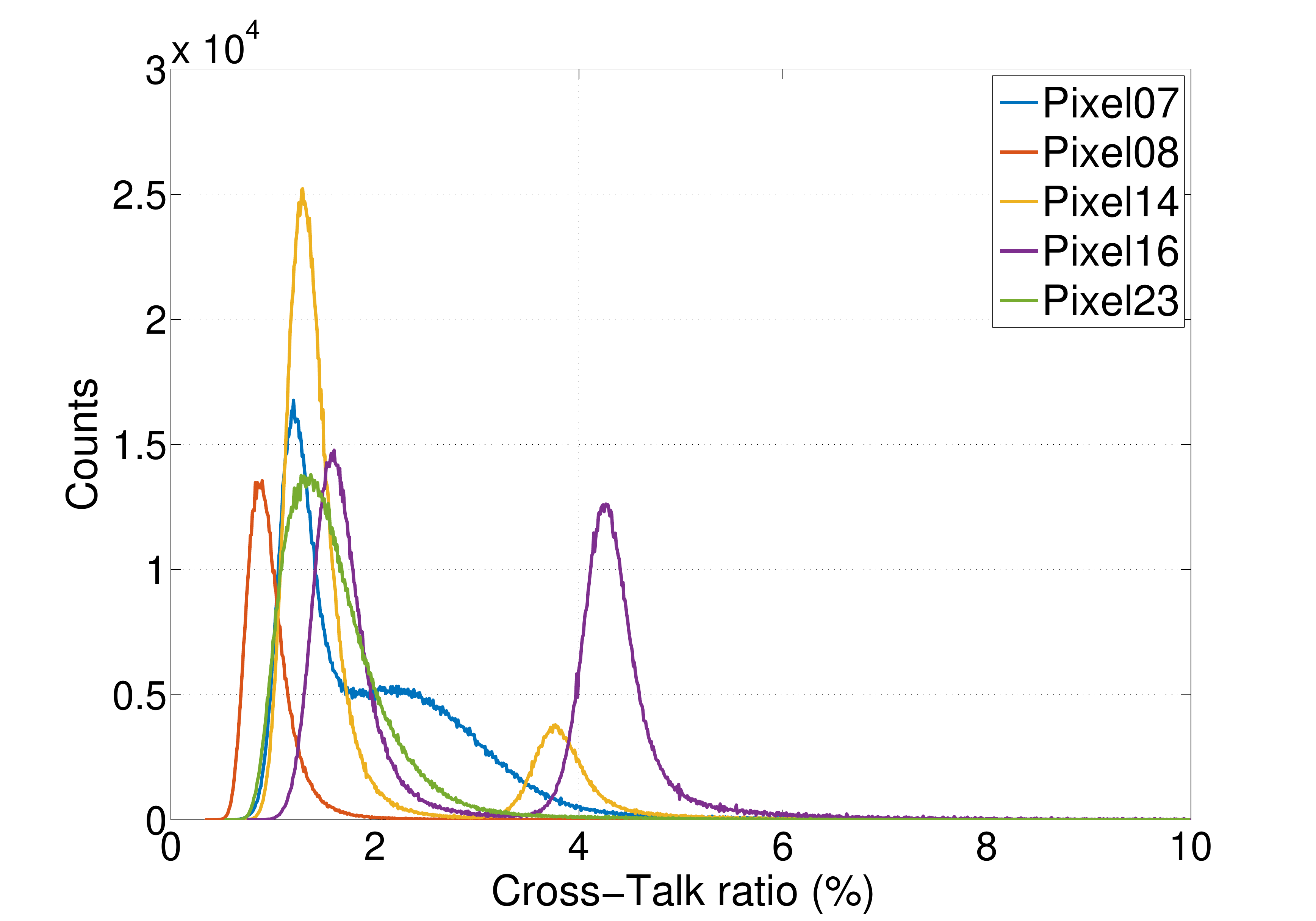}
			\caption{Cross-talk ratio distribution measured on the neighbouring channels of the uniformly illuminated pixel 15 (H12700 MaPMT, serial code: ZB0042). The socket provided by Hamamatsu is unsoldered and the biasing voltage is provided with a custom-made passive voltage divider (standard voltage ratio).}
			\label{fig:CrossTalkNoSocket}
		\end{minipage}
\end{figure}

As mentioned above, the stray capacitance of the electrical links  driving the signal from the anodes to the read-out electronics has to be minimized as it contributes to the CT$_{coupling}$ effect. In this regard, the socket embedded in the H12700 MaPMT could also add a non-negligible effect. In order to highlight this contribution, pixel 15 of a tube (serial code: ZB0042) was uniformly illuminated and the cross-talk ratio distribution was acquired in the neighbouring pixels with the usual setup. The obtained results are shown in fig.\ref{fig:CrossTalkSocket}. 

While the lower amplitude peak of pixel 7, 14 and 23 are centered at $\sim1\%$, higher values ($\sim3\%$) are observed in pixels 8 and 16, located close to the biasing high voltage pins. Later, the socket equipping the device was unsoldered and the measurements were performed on the same pixels keeping unchanged the setup. This second cross-talk amplitude distribution is reported in fig.\ref{fig:CrossTalkNoSocket}. It can be observed that the distributions of the pixels not contiguous to the biasing pins remains almost unchanged. Moreover, the lower amplitude peak of pixels 8 and 16 are now centered at $\sim 1.5\%$, similar to all the other tested pixels. This suggests that the socket standardly provided by Hamamatsu is responsible of an extra capacitive coupling to the biasing nodes which doubles the bipolar shaped cross-talk signal amplitude of the pixels located near those pins. 

Besides cross-talk, there is another known mechanism that could induce spurious signals in photomultipliers, by generating delayed afterpulses following the main signal. 
According to \cite{HamamatsuHandbook}, MaPMTs can be affected by two types of afterpulses. The first occurs due to the elastic scattering electrons produced at the first dynode which cause small amplitude signals following the main one with a very short delay (several nanoseconds to several tens of nanoseconds). In our measurements, the anode current is integrated with a time constant of the same order and an acquisition window which lasts $\sim 500$ ns  per signal (see fig.\ref{Fig:CrossWaveforms}). Therefore, this class of afterpulses is triggered together with the corresponding main signal without significantly affecting the amplitude of the signal of interest, causing marginal effects.
A second category of afterpulses is due to ion feedback which induces high amplitude signals delayed by several hundreds of nanoseconds to over a few microseconds. This kind of afterpulses is expected to be larger than single photon signals so that these two contributions can be easily separated. We treated them similar to high amplitude dark events, since we were not able to discriminate these two effects.
Although our setup (based on single photons emitted at random by a DC-biased LED) is not able to quantify the afterpulse probability for the present device, it is expected to be below 5\% for a PMT with a good quality vacuum, as explained in \cite{Afterpulses}.

%%%%%%%%%%%%%%%%%%%%%%%%%%%%%%%%%%%%%%%%%%%%%%%%%%%%%%%%%%%%%%%%%%%%%%%%%%%%%%%%%%%%%%%%%%%%%%%%%%%%%%%%%%%%%

\section{Behaviour in critical environment condition} \label{sec:Environment}
As mentioned, the features of the H12700 and the R12699 MaPMTs make them tailored for an application in high energy physics, such as in the Ring Imaging Cherenkov (RICH) detectors. %In particular, these devices are planned to be employed in the peripheral areas of the upgraded LHCb RICH-2 detector devoted to the identification of high-momentum particles. 
The magnetic field usually present in these detectors could significantly affect the photodetector performance making it important to characterize the tubes with respect to this critical environment. In addition to that, the MaPMTs are sensitive to the operating temperature which can alter the gain and cause a dark current increase. In the case of LHCb RICH upgrade, the photosensor is expected to withstand the intense illumination of up to $\sim10^6$ photons per pixel per second in the most illuminated detector areas. %which will reach a proton-proton collision rate of 40~MHz. %which will reach a proton-proton collision rate of 40~MHz. 
In the next paragraphs, the investigations about the behaviour of the H12700 and R12699 MaPMTs as a function of a longitudinal magnetic field up to 100~G (sect. \ref{sec:MagneticField}), different operating temperatures (sect. \ref{sec:Temperature}) and after a long period of light illumination (sect. \ref{sec:Aging}) are presented.

\subsection{Behaviour in magnetic field}\label{sec:MagneticField}

\begin{figure}[h!]
	\centering
		\includegraphics[width=0.45\textwidth]{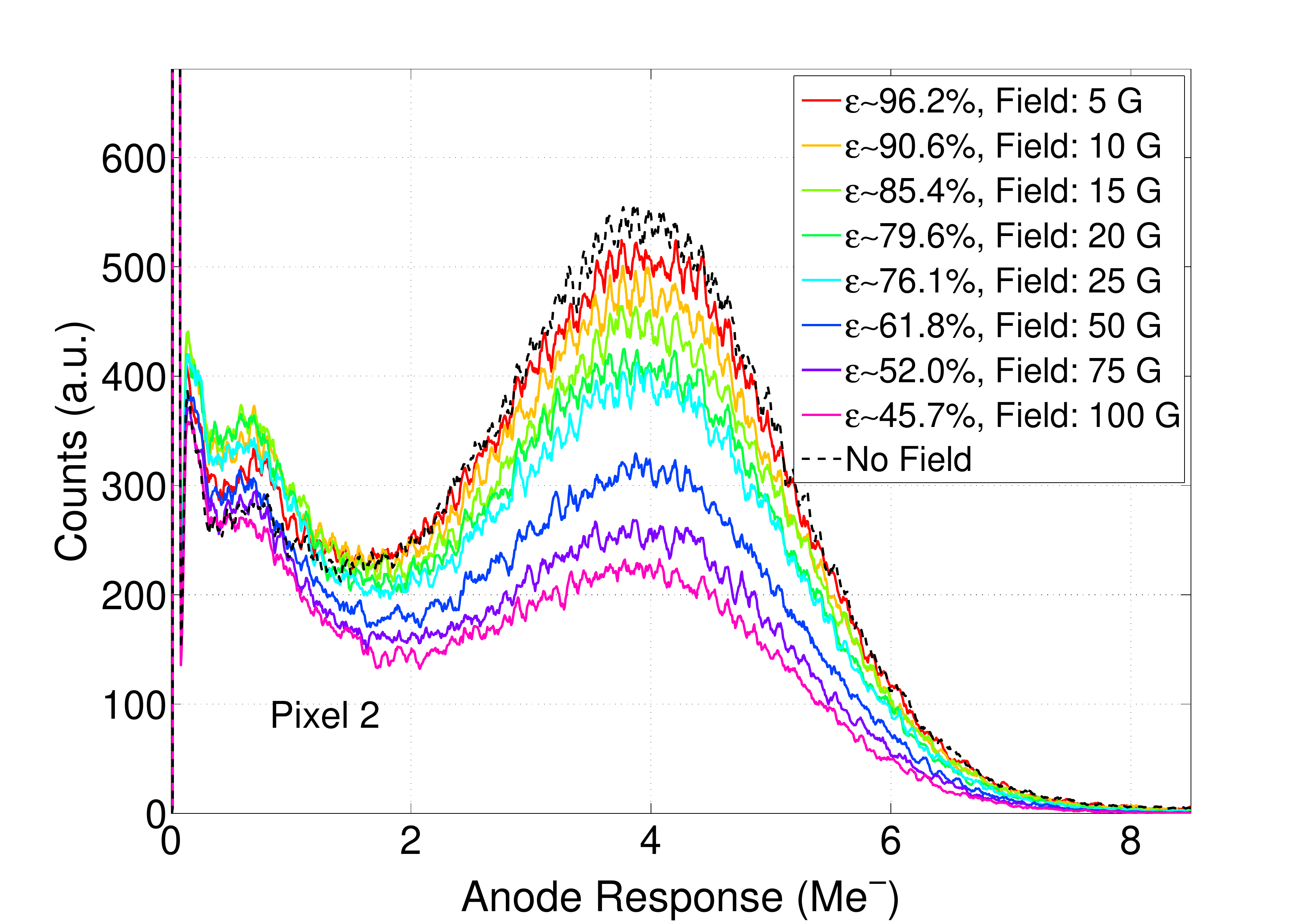}
		\hspace{5mm}%
		\includegraphics[width=0.45\textwidth]{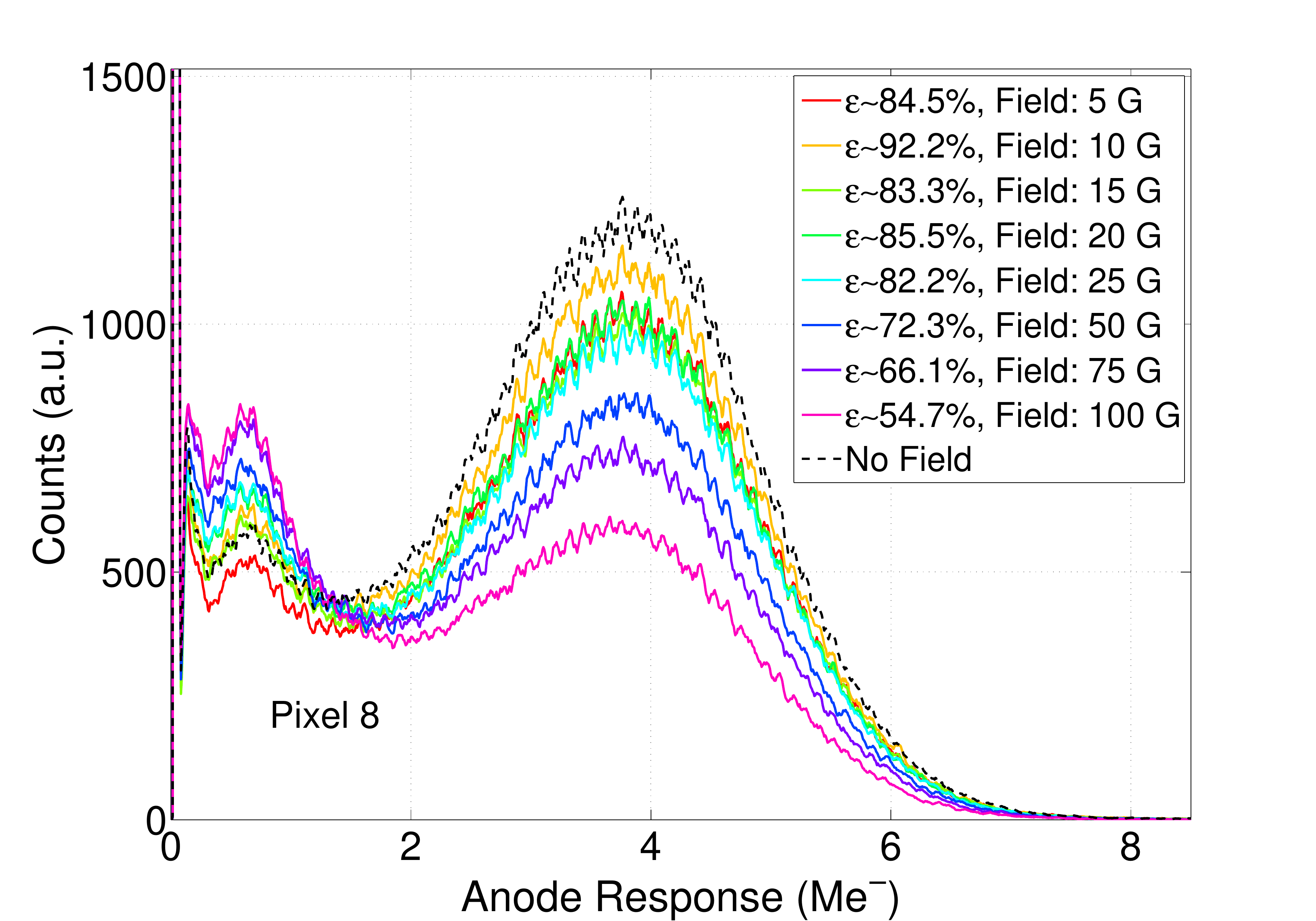}
		\includegraphics[width=0.45\textwidth]{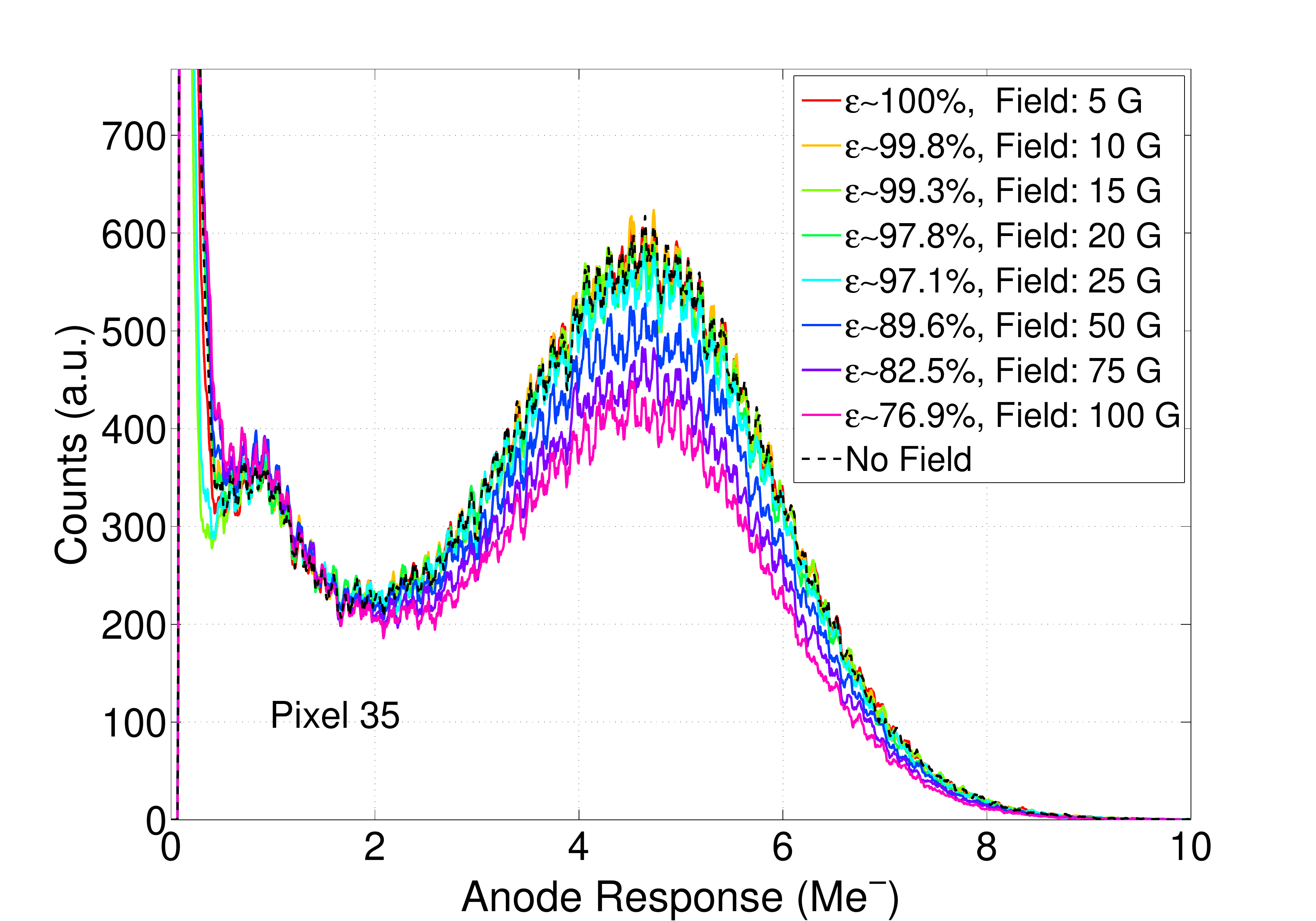}
		\hspace{5mm}%
		\includegraphics[width=0.45\textwidth]{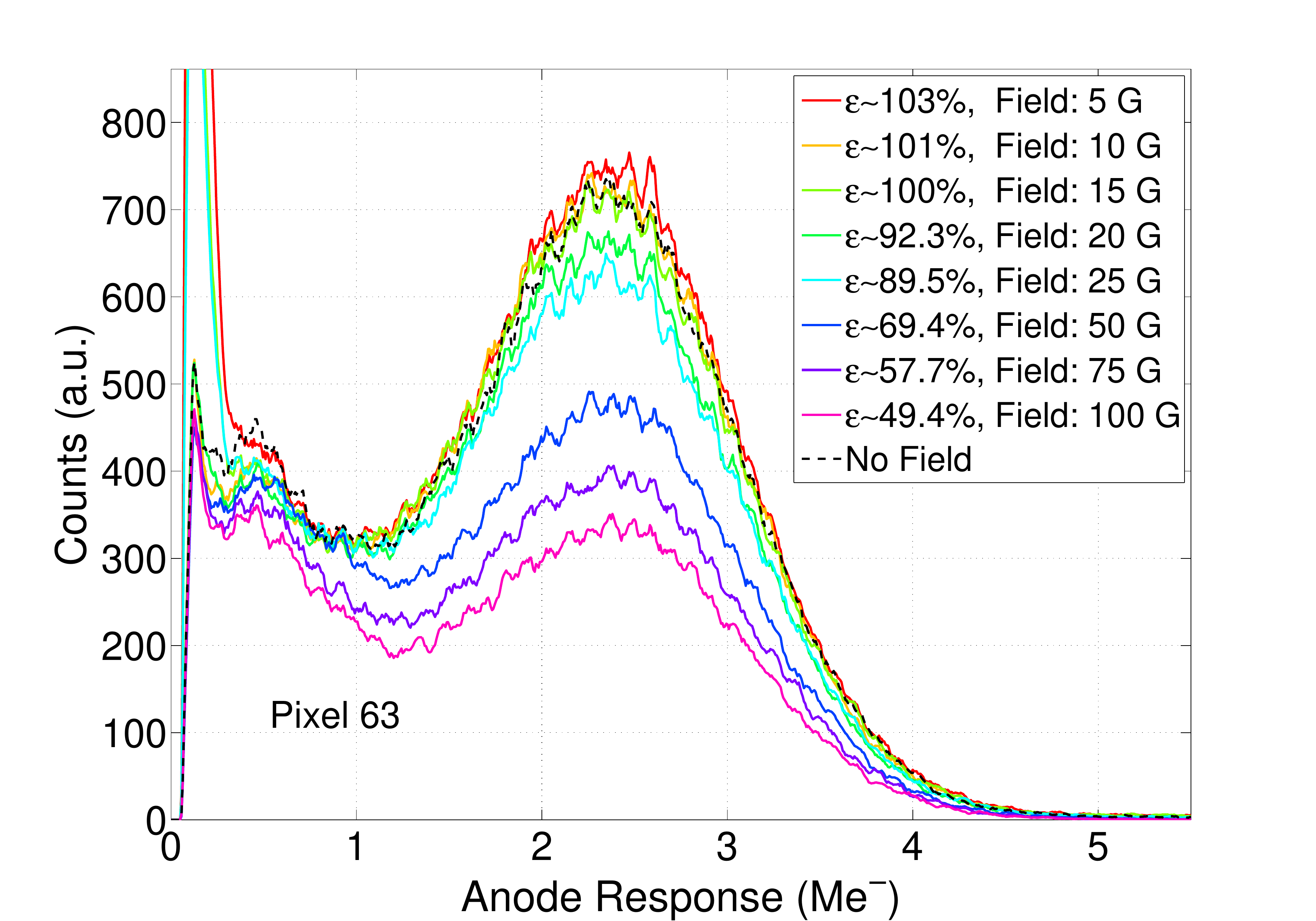}
		\caption{Superposition of single photon spectra acquired under the action of different longitudinal magnetic fields. The top plots and the bottom right one refer to pixels 2, 8 and  63 respectively, located in the external region of the MaPMT, near the biasing high voltage pins. The bottom left plot shows the results in the central pixel 35. No high magnetic permeability material was used to shield the photon detector (serial code: ZB0030, HV$=1050$~V).}
	\label{fig:NoShield}
\end{figure}

%The \itshape H12700 \upshape is supposed to be used in the peripheral regions of the Upgraded LHCb RICH-2 detector and it has to operate under the action of the fringe static magnetic field. 
In principle the performance of a photomultiplier tube is affected by an external magnetic field since it might induce some electrons changing their trajectory starting from the photocathode. %, moving from the photocathode to the first dynode or from one dynode to the following one, changing their trajectory. %According to \cite{bib3}, the highest magnetic field expected in the RICH-2 detector amounts to $\leq 15$ G. 
According to the results already obtained studying the R7600 \cite{bib6} and the R11265 \cite{NostroArticolo} MaPMTs, the main effects are caused by a longitudinal magnetic field (parallel to the tube axis and perpendicular to the photocathode surface). In a very conservative approach, the tube performance was tested under the action of a magnetic field up to 100 G produced by a solenoid. As usual, the single photons were produced using a commercial blue LED and the output signals were amplified with a charge sensitive preamplifier and acquired by a CAEN Desktop Digitizer. Figure \ref{fig:NoShield} shows the single photon spectra observed on various pixels at different magnetic field intensities.

%Senza Shield

The central pixel 35 is quite insensitive to the external magnetic field since significant degradation of the spectra is visible only for fields higher than 25 G. Defining the efficiency $\epsilon$ as the number of events with an amplitude larger than 1~$\mathrm{Me^-}$ divided by the one triggered above the same threshold without the magnetic field, the loss of efficiency at 25~G is almost negligible ($\sim 3$\%). % Thus, such central pixels would behave adequately even in the presence of longitudinal magnetic field twice as large as that expected in the RICH-2 detector. 
On the contrary, the pixels located in the peripheral area of the photon detector and near the HV pins turned out to be more sensitive. 
As fig.\ref{fig:NoShield} shows, the loss of efficiency at 25~G measured on pixel 63 is more than 10\%. An even more critical behaviour is visible in pixels 1 to 8. In this case, the loss of efficiency rises to 20-25\% at 25~G. The number of events in the single photon peak decreases significantly even for weaker fields. 

As expected, the events affected by a non-ideal electron path along the multiplication chain happen more frequently if an external magnetic field is applied. This effect is clearly visible from the number of events at $\sim0.7$~$\mathrm{Me^-}$, which increases with the field intensity. 

In order to recover the nominal performance of these peripheral pixels, the photodetector was wrapped in a high magnetic permeability material to absorb the field. In particular, the performance of the H12700 was studied in presence of a longitudinal magnetic field up to 100~G while shielding the device with one layer of Skudotech\footnote{~http://www.seliteskudotech.it} ($\sim200$~$\mathrm{\mu m}$ thick, nominal maximum magnetic permeability larger than $10^5$, produced by SELITE. Skudotech has magnetic perfomances very close to those of MUMETAL, but is much more malleable) wrapping the MaPMT lateral surface.

%Con Shield
\begin{figure}[h!]
	\centering
		\includegraphics[width=0.45\textwidth]{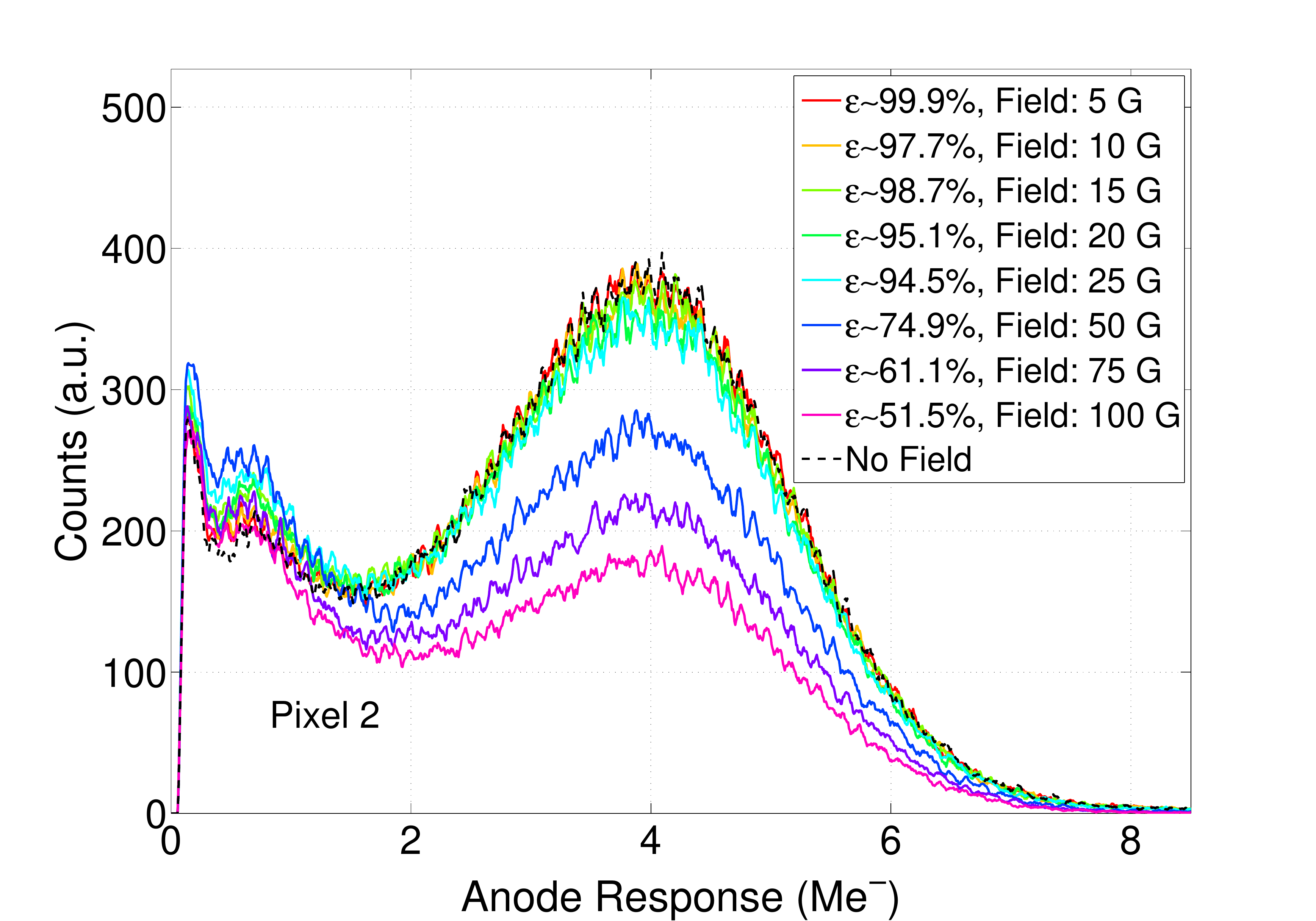}
		\hspace{5mm}%
		\includegraphics[width=0.45\textwidth]{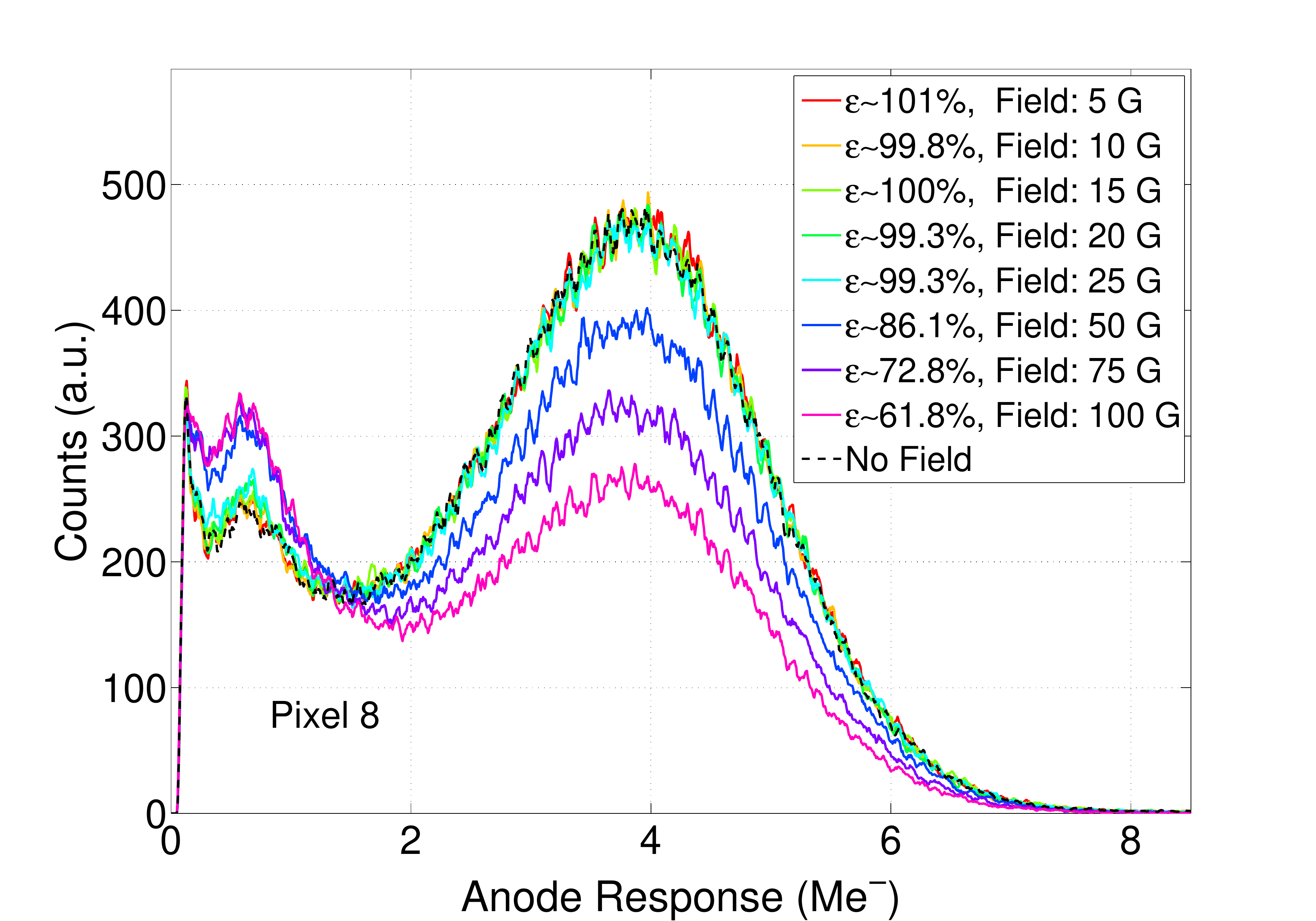}
		\caption{Superposition of single photon spectra acquired under the action of different longitudinal magnetic fields. A single Skudotech layer (thickness $\sim 200$ $\mathrm{\mu m}$) wraps the H12700. The left and right plots refer to pixels 2 and 8 respectively, which represent the worst case condition since they are located near the biasing high voltage pins  (serial code: ZB0030, HV$=1050$~V).}
	\label{fig:Shield}
\end{figure}

The shield protrudes by 1~cm from the photocathode surface, while the anode pins were kept uncovered. Figure \ref{fig:Shield} shows the results of these measurements for two of the most sensitive pixels. It can be noticed that the efficiency is almost fully recovered at moderate magnetic field intensities ($\leq 25$ G), and the loss of efficiencies reduces at most to $\sim 5$\%. Also the spectrum distortions previously observed (see fig.\ref{fig:NoShield}) are recovered if a magnetic shield is used. In presence of fields larger than 25 G, a single Skudotech layer saturates and loses its efficiency. %As far as the LHCb RICH-2 detector is concerned, the shield thickness tested here should ensure an adequate absorption for fields even two times larger than those expected.

%%%%%%%%%%%%%%%%%%%%%%%%%%%%%%%%%%%%%%%%%%%%%%%%%%%%%%%%%%%%%%%%%%%%%%%%%%%%%%%%%%%%%%%%%%%%%%%%%%%%%%%%%%%%%

\subsection{Temperature dependence}\label{sec:Temperature}
%In order to better predict the behaviour of the MaPMTs under investigation in environmental conditions similar to those expected in the Upgraded LHCb RICH-2 detector, the effects of a temperature variation need to be studied. 
The behaviour of the MaPMTs under investigation was studied with respect to the effects of a temperature variation. Indeed, increasing the operating temperature, the MaPMT gain decrease while spurious dark counts get more frequent. The setup to perform these measurements was the following: the MaPMT and the read-out electronic circuitries were put in a climatic chamber (Votsch VC 4018) operating at a temperature ranging from $10\,^{\circ}\mathrm{C}$ to $50\,^{\circ}\mathrm{C}$. The behaviour of the electronic read-out chain as a function of the temperature was previously characterized so that its contribution could be subtracted from the one of the photodetector. The single photon signal was provided by a commercial blue LED located outside of the climatic chamber and operating at constant temperature. 

\begin{figure}[h!]
	\centering
		\begin{minipage}[t]{.4825\textwidth}
			\centering
			\includegraphics[width=1\textwidth]{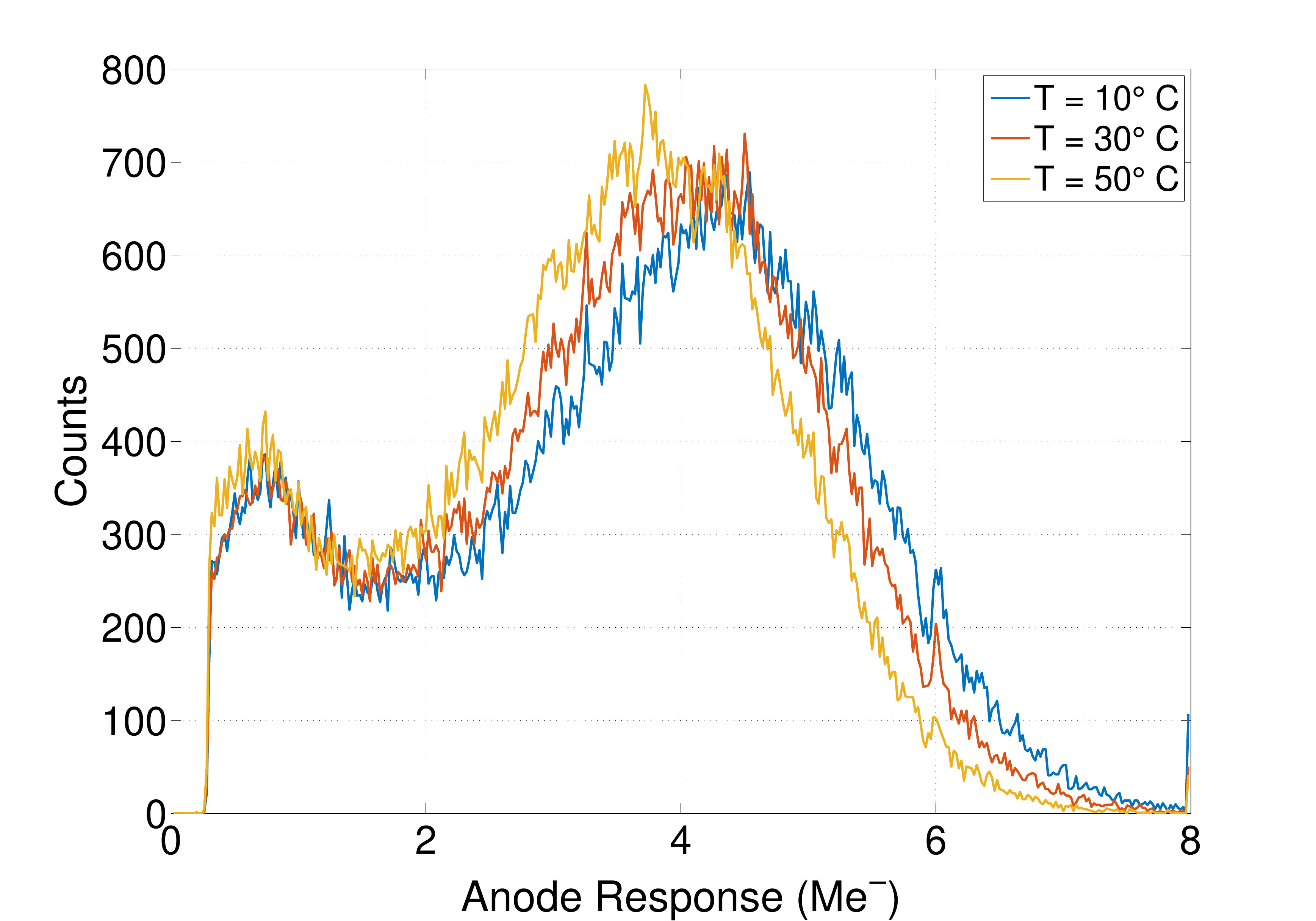}
			\caption{Single photon spectra as a function of temperature (H12700 MaPMT, serial code: ZB0030, Pixel 15,  HV$=1050$~V). The gain variation can be estimated from the single photon peak position.}
			\label{fig:GainTemperature}
		\end{minipage}%
	\hspace{5mm}%
		\begin{minipage}[t]{.4825\textwidth}
			\includegraphics[width=1\textwidth]{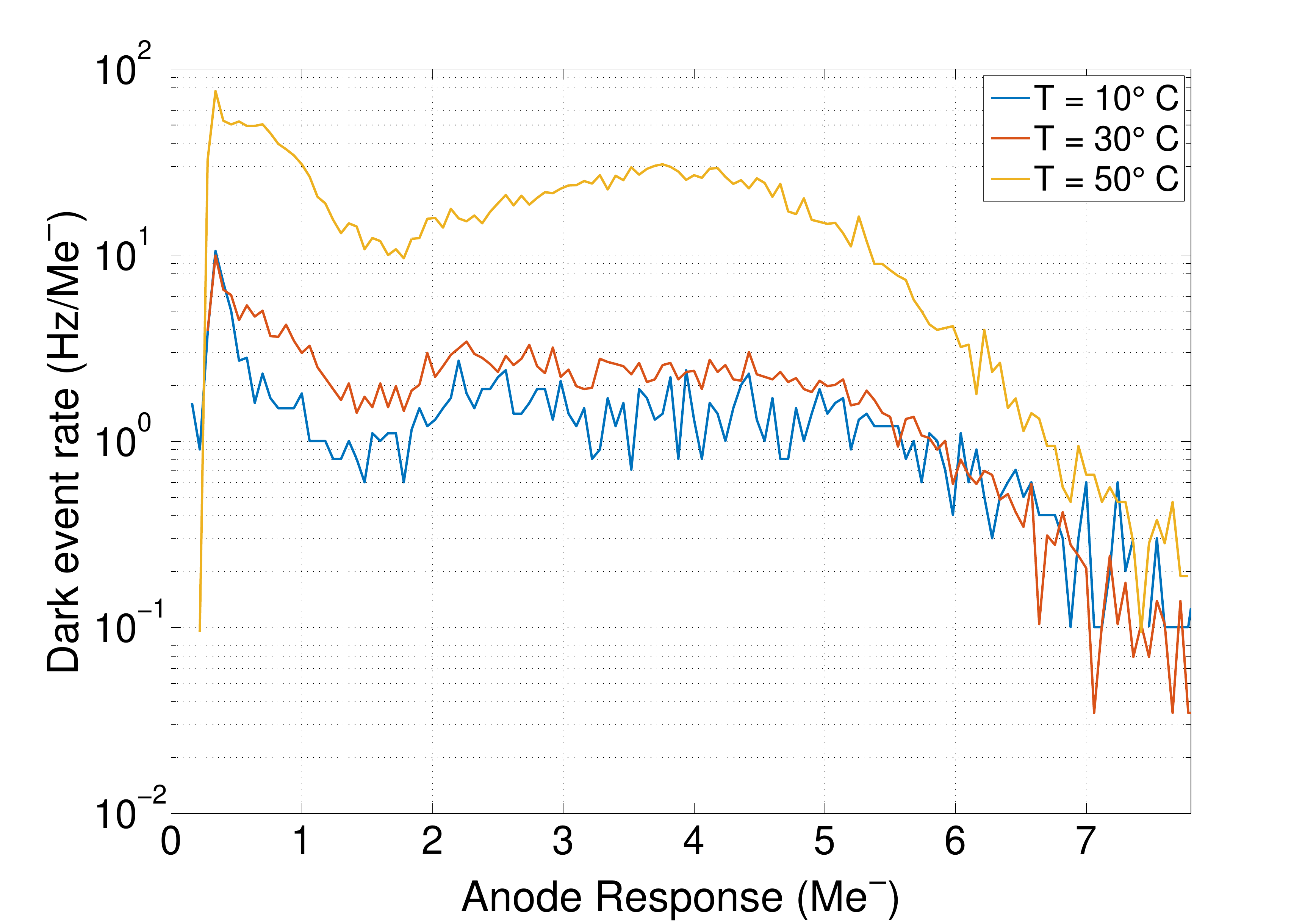}
			\caption{Dark rate spectra, in logarithmic scale, as a function of temperature (H12700 MaPMT, serial code: ZB0030, Pixel 15,  HV$=1050$~V).}
			\label{fig:NoiseTemperature}
		\end{minipage}
\end{figure}

Figure \ref{fig:GainTemperature} shows the superposition of single photon spectra acquired in the same pixel at different temperature. As it can be estimated from the single photon peak position, the gain reduces with increasing the temperature. This behaviour is well known and studied (\cite{bibt1}, \cite{bibt2}, \cite{bibt3}) and has also been observed in the R11265 tubes (\cite{NostroArticolo}). It can be explained as follows: the energy of any electron during the multiplication chain does not change with temperature as it is determined only by the electric field between dynodes. Thus, keeping the biasing voltage constant, the electrons hit a certain dynode with similar energies and predominantly interact with it by electron-electron interactions. The amount of charges released by these interactions does not depend on temperature but mainly on the energy of the primary electron. The secondary electrons generated have a low kinetic energy so that they manly interact with phonons. This interaction has a strong temperature dependence so that the mean free path of electrons inside the dynode decreases with increasing temperature. Therefore, although a fixed amount of free charges is generated, a smaller number of electrons is able to escape from the dynode surface and take part in following multiplication steps. The global result is that the mean number of secondary electrons emitted from each dynode reduces, with a consequent gain decrease. 
From fig.\ref{fig:GainTemperature} it is possible to estimate that the gain reduction as a function of the temperature is quite linear in the range of interest and the characteristic slope amounts to about -10 ke$^-/^{\circ}\mathrm{C}$ or 0.25 $\%/^{\circ}$C.

Another parameter that strongly depends on temperature is the dark current. Indeed, the rate of noise signals increases greatly with temperature as shown in fig.\ref{fig:NoiseTemperature} where dark counts spectra acquired at different temperature are superimposed with a logarithmic scale. The reason of this phenomenon is that, increasing the temperature, the number of electrons which have enough thermal energy to escape from the dynodes or the photocathode surface and generate a multiplication process increases. In particular, the dark counts rate at 50$^{\circ}$ C above 1 Me$^-$ is more than an order of magnitude larger than that at 10$^{\circ}$ C.

%%%%%%%%%%%%%%%%%%%%%%%%%%%%%%%%%%%%%%%%%%%%%%%%%%%%%%%%%%%%%%%%%%%%%%%%%%%%%%%%%%%%%%%%%%%%%%%%%%%%%%%%%%%%%

\subsection{Aging Test}\label{sec:Aging}

Long periods of light exposure cause MaPMT aging and will therefore deteriorate the device's initial performance. This critical condition is common in high energy physics applications where the device must withstand extremely high photon rates for thousands of hours. Typical effects due to the aging of the device are the reduction of the photocathode efficiency, the dark count rate increase and the variation of the gain of the tube. From the experience gained during the characterization of the R11265-103-M64 MaPMT \cite{NostroArticolo}, the most critical effect due to the aging is the gain loss caused by the deterioration of the dynodes which change the mean number of secondary electrons emitted. 

In order to make a quantitative estimation of how much the gain is affected by aging, a fully automatic system was set up to age a H12700 and a R12699 MaPMTs. A commercial blue LED illuminated the device under test, causing a stable aging current. Periodically, the LED biasing voltage was turned down so that it operated in single photon regime. The single photon signals from the tested pixels were amplified by charge sensitive amplifiers and acquired with three DT5720 Desktop Digitizers (CAEN). The gain loss was evaluated by measuring the single photon peak position over several thousand hours of effective LED illumination. A slow control system continuously monitored the DC aging current in all the device and ensured the temperature stability by adjusting the heat injected by four power resistors. This prevents the temperature variation to affect the gain of the photosensor during the measurement. Further details on the setup conditions and the analysis can be found in \cite{NostroArticolo}.

\begin{figure}[h!]
	\centering
		\begin{minipage}[t]{.4825\textwidth}
			\centering
			\includegraphics[width=1\textwidth]{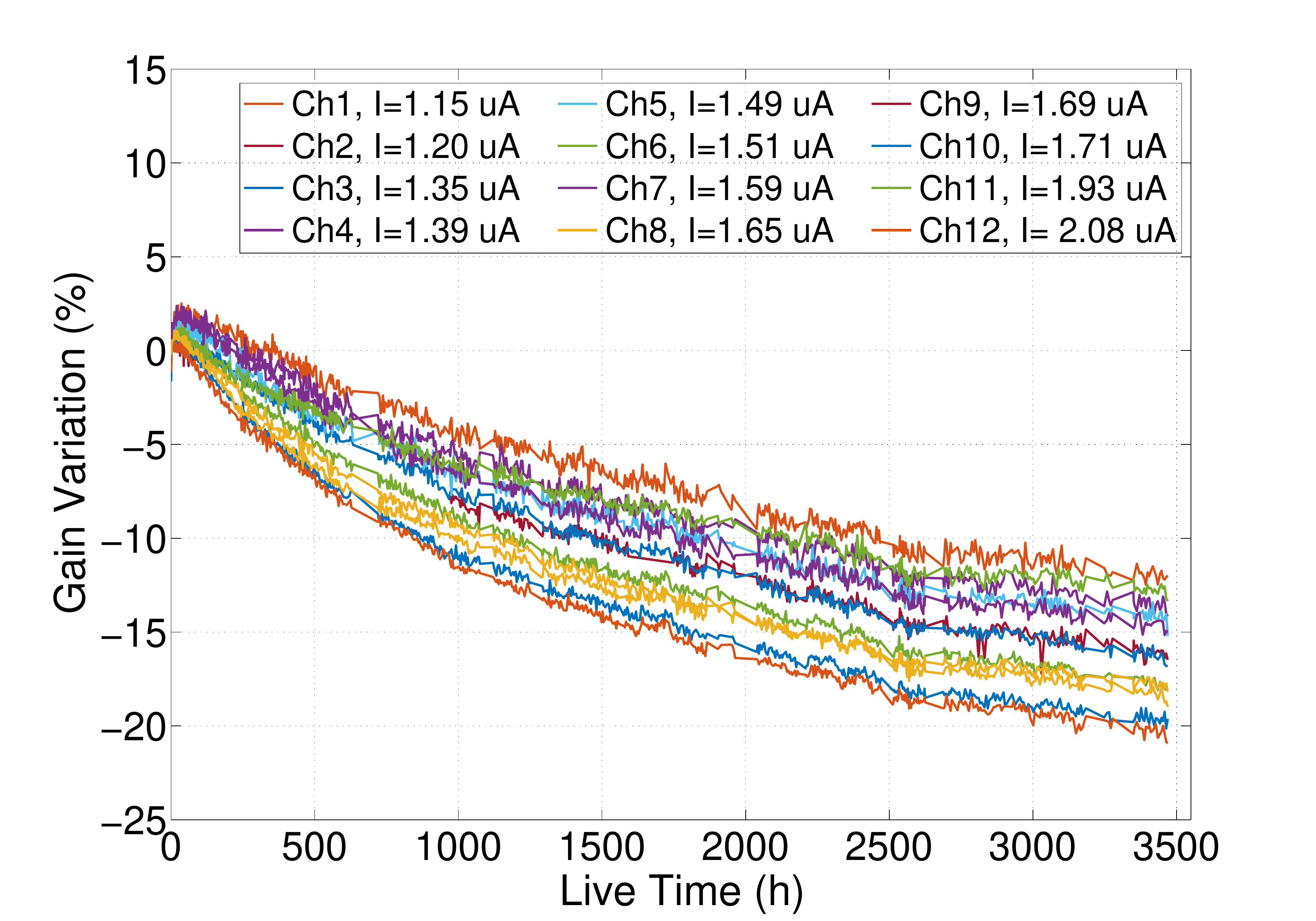}
			\caption{Gain variation (in \%) versus the illumination period for the tested H12700 MaPMT (serial code: ZB0030, HV$=1000$~V, T$=24.5^{\circ}$ C).}
			\label{fig:Aging_H12700}
		\end{minipage}%
	\hspace{5mm}%
		\begin{minipage}[t]{.4825\textwidth}
			\includegraphics[width=1\textwidth]{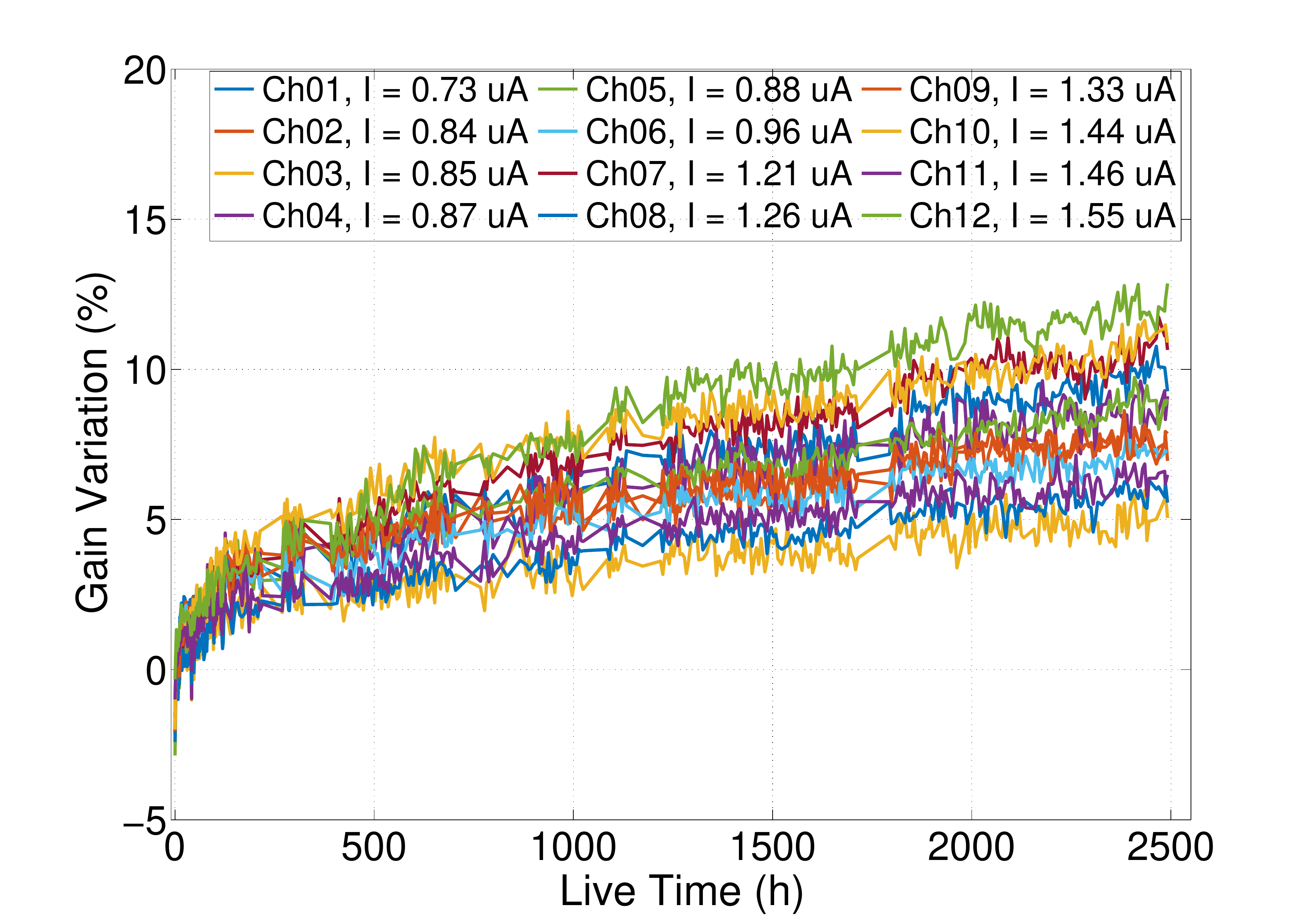}
			\caption{Gain variation (in \%) versus the illumination period for the tested R12699 MaPMT (serial code: HA0016, HV$=900$~V, T$=25^{\circ}$ C). }
			\label{fig:Aging_R12699}
		\end{minipage}
\end{figure}

As mentioned above, two MaPMTs have been tested (one H12700, serial code: ZB0030, and one R12699, serial code: HA0016) and the gain variation as a function of the effective LED illumination period are shown in fig.\ref{fig:Aging_H12700} and fig.\ref{fig:Aging_R12699}. The gain loss is calculated as the percentage variation of the single photon peak position with respect to the value measured at the beginning in each pixel, after keeping the device in the dark for 24 hours and illuminating it for one hour. 

Referring to fig.\ref{fig:Aging_H12700}, the illumination level was set in order to induce an initial integrated current over the anodes of $\simeq 94$ $\mu$A, similar to the maximum value sustainable by the tube according to the manufacturer specifications (100 $\mu$A). Such an intense illumination, roughly equivalent to a single photon rate of few MHz per pixel, was chosen because %it represents the very worst case condition expected in the LHCb RICH detector environment and also because 
it is similar to the current value standardly used in this test by the manufacturer. All channels are observed to have a similar trend and the gain loss after 3500 hours ranges from $\sim 13$\% down to $\sim 20$\%. %The gain loss only slightly depends on the DC aging current as the slope of the curves are similar and the difference between the pixel behaviours is dominated by the initial offset. 
The slope of the curves gradually decreases with the LED illumination period and becomes almost flat above $\sim 2800$ hours.  In particular, the gain loss in the last $\sim 700$ hours of LED illumination is of the order of 1\% and comparable with the measurement uncertainties. In this plateau region the performance of the photodetector can be assumed stable. 

Figure \ref{fig:Aging_R12699} shows the result of the same measurement performed on the R12699 MaPMT, serial code HA0016. The device was biased at HV$=900$~V with a consequent reduction of the initial anode current to $\sim40$ $\mu$A. The absolute gain variation decreases with respect to the one observed in the former tube. This suggests that the gain variation is related to the integrated DC current which is responsible of the deterioration of the dynodes and depends on the illumination level and the average gain of the device.

What stands out comparing the two measurements is that the two photodectors show opposite trends. Indeed, the gain of all the observed pixels of fig.\ref{fig:Aging_R12699} increases with the illumination period in contrast with what observed in the other device (fig.\ref{fig:Aging_H12700}). This strange behaviour was already exhibited by some R11265 samples, as presented in \cite{NotaPubblicaCERN}. 
Hamamatsu ascribes this to the spread of the thickness of the Cesium layer grown on the dynode surfaces \cite{Cesium}. %An optimized level of Cesium increases the dynode efficiency since it lets the secondary electrons easily overcome the potential barrier at the interface between the dynode surface and the vacuum, promoting their emission via tunnel effect. Therefore, growing a proper level of Cesium ensures higher gain. Nevertheless, if the Cesium layer is thicker than the optimal value, then the probability of electron emission via tunnel effect decreases. Long period of LED illumination causes that the signal electrons wear the Cesium away from the dynodes surface with a consequent variation of the device gain. Now, if the starting Cesium layer was equal to the optimal thickness, the aging would result in a gain reduction. On the other hand, if the initial Cesium layer was thicker than the optimal, its erosion would lead toward an optimization and a gain increase. Since the thickness of that layer is a parameter hard to be controlled during the production process, it is difficult to predict a priori the gain variation sign of a specific device, but photodetectors belonging to the same production batch are expected to behave similarly. 

According to our observations, the absolute gain variation amounts at worst to $\mid \Delta G \mid \simeq 20 \%$ after 3000 hours of high intense illumination (equivalent single photon rate of few MHz per pixel). Finally, note that, as the device reaches gain stability, its overall variation can be compensated by adjusting the high voltage biasing the MaPMT, as shown in \cite{NostroArticolo}.

%%%%%%%%%%%%%%%%%%%%%%%%%%%%%%%%%%%%%%%%%%%%%%%%%%%%%%%%%%%%%%%%%%%%%%%%%%%%%%%%%%%%%%%%%%%%%%%%%%%%%%%%%%%%%

\section{Conclusions}

The H12700 multi-anode photomultiplier tube has been fully characterized. It is a 64-channel square photodetector produced by Hamamatsu which is sensitive to single photon signals. Five devices have been tested, half equipped with a socket provided by the manufacturer (H12700) and half without it (R12699). The photodetector proves its capability to detect single photons with a typical gain of several Me$^-$/photon at HV$=1$ kV, with an uniformity factor of $\sim$2-3 and an extremely low dark current ($\leq 40$ $\mathrm{Hz/cm^2}$ at room temperature above 1 Me$^-$). The cross-talk was investigated in depth: the worst case condition tuned out to be due to the charge sharing among neigbouring pixels, which causes signals with an amplitude up to $\sim7 \%$ of the inducing single photon signal. The R12699 MaPMT significantly reduces the coupling between pixels located near the HV pins, thus decreasing the cross-talk. The features of these devices make them suitable for an application in high energy physics, such as in RICH detectors. The photodetectors performance have been studied with respect to a longitudinal magnetic field up to $100$ G, typical of such environments: the most critical pixels are those in the outer regions, but their proper operation can be recovered using local high magnetic permeability shields. Ranging from 10$^{\circ}$ C to 50$^{\circ}$ C, the gain of the device linearly decreases by 0.25 $\%/^{\circ}$C and the dark counts rate increases by a factor 10 in this temperature range. The gain variation due to the device aging was also characterized on two tubes. Although it is hard to define a typical behaviour since the gain may show both a positive or a negative variation, similar trends would be expected for devices of the same production batch. After 3000 hours of intense illumination, the gain variation amounts to  $\mid \Delta G \mid \simeq 20 \%$ and could be compensated by adjusting the biasing voltage as the device reaches the gain stability.

%%%%%%%%%%%%%%%%%%%%%%%%%%%%%%%%%%%%%%%%%%%%%%%%%%%%%%%%%%%%%%%%%%%%%%%%%%%%%%%%%%%%%%%%%%%%%%%%%%%%%%%%%%%%%

\end{document}